\documentclass[
reprint,
superscriptaddress,
preprintnumbers,
amsmath,amssymb,
prd,
floatfix,
showkeys
]{revtex4-2}

\usepackage{dcolumn}
\usepackage{bm}
\usepackage{natbib}
\usepackage{graphicx}
\usepackage{amsmath}
\usepackage{amssymb}
\usepackage{xcolor}
\usepackage{chngpage}
\usepackage{comment}

\usepackage{hyperref}
\usepackage{cleveref}

\renewcommand{\vec}[1]{\boldsymbol{#1}}

\newcommand{\ignore}[1]{}

\begin{document}

\preprint{FERMILAB-PUB-23-631-T}
\title{The Operator Product  Expansion  for Radial Lattice
Quantization of 3D \texorpdfstring{$\phi^4$}{p4} Theory} 

\author{Venkitesh Ayyar}
\affiliation{Department of Physics and Center for Computational Science, Boston University, Boston, MA 02215, USA}
\author{Richard C. Brower}
\affiliation{Department of Physics and Center for Computational Science, Boston University, Boston, MA 02215, USA}
\author{George T. Fleming}
\email{gfleming@fnal.gov}
\affiliation{Fermi National Accelerator Laboratory, Batavia, IL 60510, USA}
\affiliation{Department of Physics, Sloane Laboratory, Yale University, New Haven, CT 06511, USA}
\author{Anna-Maria E. Glück}
\email{anna-maria.glueck@stud.uni-heidelberg.de}
\affiliation{Kirchhoff-Institut für Physik, Ruprecht-Karls-Universität Heidelberg, 69120 Heidelberg, Germany}
\affiliation{Department of Physics, Sloane Laboratory, Yale University, New Haven, CT 06511, USA}
\author{Evan K. Owen}
\affiliation{Department of Physics and Center for Computational Science, Boston University, Boston, MA 02215, USA}
\author{Timothy G. Raben}
\affiliation{Department of Physics and Astronomy, Michigan State University, East Lansing, MI 48824, USA}
\author{Chung-I Tan}
\affiliation{Brown Theoretical Physics Center and Department of Physics, Brown University,
Providence, RI 02912, USA}

\date{\today}

\begin{abstract}

At its critical point, the three-dimensional lattice Ising model is described by a conformal field theory (CFT), the 3d Ising CFT.
Instead of carrying out simulations on Euclidean lattices, we use the Quantum Finite Elements method to implement radially quantized critical $\phi^4$ theory on simplicial lattices approaching $\mathbb{R} \times S^2$. Computing the four-point function of identical scalars, we demonstrate
the power of radial quantization by the accurate determination of the scaling dimensions $\Delta_{\epsilon}$ and $\Delta_{T}$ as well as ratios of the operator product expansion (OPE) coefficients $f_{\sigma \sigma \epsilon}$ and $f_{\sigma \sigma T}$ of the first spin-0 and spin-2 primary operators $\epsilon$ and $T$ of the 3d Ising CFT. 
\end{abstract}


\maketitle

\section{\label{sec:introduction} Introduction}

Since the first discovery of critical opaqueness made by Andrews in the 19th century \cite{Andrews1869} and especially since Wilson's breakthrough formulation of the Renormalization Group framework for critical phenomena \cite{Wilson1971}, the study of systems at criticality has fascinated physicists throughout the decades. The prototype of such a system with a continuous phase transition and the workhorse in studying critical phenomena is the Ising model. Despite its simplicity, it is of great physical relevance because universality guarantees that the critical Ising model shows the same behavior as encountered for critical transitions in e.g. uniaxial magnetic systems, fluids \cite{Parola1995}, and micellar systems \cite{Anisimov1991}. 

Motivated by experimental measurements of the scaling behaviour of physical quantities near criticality, past theoretical efforts to study the critical point of the Ising model had the main goal of calculating its critical exponents. As there is no analytic solution for the Ising model in \(d\geq3\), such calculations have to be performed using approximate and computational methods, the most prominent of those including high-temperature expansions, field-theoretic methods such as the \(\varepsilon\)-expansion or non-perturbative methods based on approximate solutions of the RG-equations, as well as Monte Carlo simulations. For a review, see \cite{Pelissetto2000}. After the development of spin cluster algorithms \cite{Swendsen1987, Wolff1989}, especially the latter method combined with finite-size scaling arguments yielded precise results for the critical exponents \cite{Pelissetto2000, Hasenbusch2010}.

For comparison with experiment, it is sufficient to determine the critical exponents of the 3d Ising model. But from a theoretical point of view, the critical Ising model contains much more information. Following Polyakov's hypothesis \cite{Polyakov:1970xd}, we expect the scale invariance of a system at criticality combined with Poincaré invariance to lead this critical system to be invariant under the conformal group \(O(d+1, 1)\). Consequently, the physics of the critical 3d Ising model can be described by a conformal field theory (CFT), the 3d Ising CFT. The local operator content of such a CFT is spanned by the so-called {\em primary operators} and their descendants, which are created by acting on the primaries with translation generators. The primary operators can be assigned spins and parity and, in the case of the 3d Ising CFT, include the leading \(0^-\)-operator \(\sigma\), the leading \(0^+\)-operator \(\epsilon\) and the energy-momentum tensor \(T\) (\(2^+\)), as well as infinitely many subleading and higher-spin operators. As will be further explained in Sec.~\ref{sec:4pt}, a CFT is characterized by two sets of quantities - the {\em scaling dimensions} \(\Delta_{\mathcal{O}}\) and the {\em operator product expansion (OPE) coefficients} \(f_{\mathcal{O}_1 \mathcal{O}_2 \mathcal{O}_3}\) for different primary operators \(\mathcal{O}\) of the CFT.

For (at least some of) the scaling dimensions, there is a mapping to the critical exponents, which have been thoroughly investigated with the traditional methods reviewed in \cite{Pelissetto2000}
\begin{align}
    \eta &= 2\Delta_{\sigma}-d+2, \nonumber \\
    \nu &= 1/(d-\Delta_{\epsilon}), \nonumber \\
    \omega &= \Delta_{\epsilon'}-d.
\end{align}
The OPE coefficients, however, are largely uninvestigated with these methods. With Functional Renormalization Group methods, it was only recently possible to determine \(f_{\sigma \sigma \epsilon}\) \cite{Rose2022}. Moreover, \(f_{\sigma \sigma \epsilon}\) and \(f_{\epsilon \epsilon \epsilon}\) have been measured with traditional Monte Carlo methods on Euclidean lattices from three-point correlators~\cite{Hasenbusch2018, Herdeiro2017} and from the scaling of 2-point correlators ~\cite{Caselle2015, Costagliola2016} of the operators \(\sigma\) and \(\epsilon\). 

However, as suggested by Cardy~\cite{Cardy:1985lth} in 1985, there is, in principle, a huge advantage by replacing the traditional toroidal Euclidean lattice approximation to ${\mathbb R}^d$ with simulations on simplicial lattices convergent to the cylindrical  ${\mathbb R} \times {\mathbb S}^{d-1}$. This is an exact Weyl map of all CFT data to what is called {\em radial quantization}~\cite{Fubini:1972mf} with translations generated by the dilatation operator. The challenge~\cite{Cardy:1985lth} for $d >2$ is to define the correct lattice action on the curved manifold. Recently, the {\em Quantum Finite Elements (QFE)}~\cite{Brower:2016moq} project
has begun to address this by including lattice counter terms to improve the UV cut-off.  

QFE tests began with $\phi^4$ theory at the critical point in comparison with the minimal 2d  Ising CFT on ${\mathbb S^2}$, accurately reproducing the exact results for the \(\mathbb Z_2\)-odd scalar propagator and the scalar four-point amplitude~\cite{Brower2018} as well as examples of the $\mathbb Z_2$-even sector in the fermionic representation~\cite{PhysRevD.95.114510} on  ${\mathbb S^2}$. The next step was to apply the QFE to critical $\phi^4$ theory on $\mathbb{R} \times {\mathbb S^2}$ in order to compare to the 3d Ising CFT. The scalar propagator~\cite{Brower:2020jqj} was well represented in the continuum limit with the
scaling dimension $\Delta_\sigma$ determined to better than $10^{-2}$ agreement with the conformal bootstrap~\cite{ElShowk2012, ElShowk2014, SimmonsDuffin2016, Reehorst2022}. Here, we extend the QFE method to the scalar four-point amplitude of the 3d Ising CFT. The present work is intended as a proof of concept to show that by fitting our lattice data to the expected form from the continuum operator product expansion (OPE) for CFTs in radial quantization, we can extract scaling dimensions $\Delta_{\mathcal{O}}$ and OPE coefficients $f_{\sigma \sigma \mathcal{O}}$ of \(\mathbb Z_2\)-even primary operators in the continuum limit. Because an exact OPE is an intrinsic property of CFTs, this provides a stringent test of whether we actually reach the 3d Ising CFT in the continuum with our QFE radial quantization. Preliminary results presented at LATTICE 2022~\cite{Gluck:2023zji} are extended and finalized here to higher fidelity to assess the capabilities of this method.

In the conclusion, we discuss further improvements of the QFE method and make comparisons with other approaches. In particular, the conformal bootstrap spectacularly constrains the four-point functions, giving rigorous bounds not only on the scaling dimensions of whole sets of conformal primary operators but also on their OPE coefficients \cite{SimmonsDuffin2016, Reehorst2022}. Even for non-integral theories, these offer definitive tests for Monte Carlo simulations. In addition, there are recent successes in determining OPE coefficients using Hamiltonian methods on the fuzzy sphere \cite{Han:2023yyb, Hu:2023xak}. While each method is complementary, comparisons may suggest improvement for them all.

\section{\label{sec:4pt} \texorpdfstring{$n$}{n}-point functions in radially quantized CFTs}

\subsection{CFTs and Radial Quantization}
At the Wilson-Fisher fixed point, 3d \(\phi^4\) theory is described by the 3d Ising conformal field theory (CFT). \(d\)-dimensional CFTs are quantum field theories that are invariant under the conformal group \(O(d+1, 1)\) of transformations that leave the metric invariant up to a Weyl rescaling \cite{DiFrancesco1997}
\begin{equation}
    g_{\mu\nu}(\vec{x}) \rightarrow g_{\mu\nu}'(\vec{x'}) = \Lambda^2(\vec{x}) g_{\mu\nu}(\vec{x}).
    \label{eq:Weyl}
\end{equation}
Thus, instead of considering the 3d Ising CFT on three-dimensional Euclidean space with metric 
\begin{equation}
ds_{flat}^2 = dr^2 + r^2 d\Omega^2 = \frac{r^2}{R^2} \left[dt^2 + R^2 d\Omega^2 \right]
\end{equation}
where \(t =  R \log(r/R)\), one can quantize the theory on \(\mathbb{R}\times S^2\) with Weyl-rescaled metric 
\begin{equation}
ds_{cyl}^2 = dt^2+R^2d\Omega^2 = \frac{R^2}{r^2} ds_{flat}^2.
\end{equation}
$n$-point functions of scalar fields in flat space are then related to those on the cylinder via
\begin{align}
    & \langle \sigma(r_1, \Omega_1) \sigma(r_2, \Omega_2) \cdots \rangle_{flat} \nonumber \\ 
    & = \left(\frac{R}{r_1}\right)^{\Delta_{\sigma}} \left(\frac{R}{r_2}\right)^{\Delta_{\sigma}}  \cdots \langle \sigma(t_1, \Omega_1) \sigma(t_2, \Omega_2) \cdots\rangle_{cyl}
    \label{eq:nptrescaling}
\end{align} 
with \(\Delta_{\sigma}\) the scaling dimension of scalar primary \(\sigma\) \cite{Rychkov:2016iqz}.
This choice of coordinates is called radial quantization and is often useful for scale-invariant theories as dilatations of \(r\) on \(\mathbb{R}^3\) are mapped to shifts in cylinder time \(t\). As will be explained in subsection \ref{subsec:antipodal4pt}, the determination of CFT data is also extremely facilitated by radial quantization. 

\subsection{Conformal \texorpdfstring{$n$}{n}-point functions}
Due to the symmetries of CFTs, two-point functions of primary operators follow power laws \cite{DiFrancesco1997}
\begin{align}
    \langle \mathcal{O}(\vec{x_1}) \mathcal{O}(\vec{x_2}) \rangle = \frac{C_{\mathcal{O}}}{|\vec{x_1}-\vec{x_2}|^{2\Delta_{\mathcal{O}}}}.
\end{align}
Moreover, the operator product expansion (OPE) is exact, such that under the path integral, a product of two primaries can be replaced by a sum over ``intermediate states'' (schematically)
\begin{align}
    \mathcal{O}_i(x_1)\mathcal{O}_j(x_2) = \sum_{k} f_{\mathcal{O}_i \mathcal{O}_j \mathcal{O}_k}  C(x_1-x_2) \mathcal{O}_k(x_2),
    \label{eq:OPEschematic}
\end{align}
with the OPE coefficients \(f_{\mathcal{O}_i \mathcal{O}_j \mathcal{O}_k}\) as expansion coefficients. Thus, all $n$-point functions of a CFT can be expressed as functions of the scaling dimensions \(\Delta_{\mathcal{O}}\) and the OPE coefficients \(f_{\mathcal{O}_i \mathcal{O}_j \mathcal{O}_k}\) of conformal primaries \(\mathcal{O}\). Moreover, all those quantities can be extracted by measuring different four-point functions.  

In this work, we focus on the four-point function of four identical scalar primaries \(\sigma\), which - when divided by the corresponding two-point functions - forms the amplitude \(g(u, v)\). It is called the conformally invariant amplitude as any factors~Eq.\eqref{eq:nptrescaling} from Weyl rescaling the metric~Eq.\eqref{eq:Weyl} drop out. Thanks to the OPE, this amplitude can be expanded into a sum over the primaries \(\mathcal{O}\) of the CFT \cite{Dolan:2000ut, Dolan:2011dv}: 
\begin{align}
g(u,v) &\equiv \frac{\langle \sigma(x_1)\sigma(x_2)\sigma(x_3)\sigma(x_4)\rangle}{\langle \sigma(x_1)\sigma(x_2) \rangle \langle \sigma(x_3)\sigma(x_4) \rangle} \label{eq:guv} \\
&= 1 + \sum_{\mathcal{O}}  f_{\sigma \sigma \mathcal{O}}^2 G_{\mathcal{O}}(\Delta_{\mathcal{O}}; u, v).
\label{eq:OPEexp} 
\end{align}
In this case, actually only the primaries \(\mathcal{O}\) with even spin and parity contribute because of the symmetries of the amplitude. The \(G_{\mathcal{O}}\) are the so-called conformal blocks, functions of the scaling dimensions \(\Delta_{\mathcal{O}}\) of primaries \(\mathcal{O}\) as well as the kinematic information encoded in the conformally invariant cross ratios \(u\) and \(v\)
\begin{align}
    \sqrt{u} = \frac{|\vec{x_1}-\vec{x_2}||\vec{x_3}-\vec{x_4}|}{|\vec{x_1}-\vec{x_3}||\vec{x_2}-\vec{x_4}|},  \  
    \sqrt{v} = \frac{|\vec{x_1}-\vec{x_4}||\vec{x_2}-\vec{x_3}|}{|\vec{x_1}-\vec{x_3}||\vec{x_2}-\vec{x_4}|}.
\end{align}

In odd dimensions, there is no known analytic form of the conformal blocks. However, if we express the \(G_{\mathcal{O}}\) as a function of the (also conformally invariant) quantities 
\begin{equation}
    \cosh(\tau) = \frac{1+\sqrt{v}}{\sqrt{u}}, \ \ \ \ \cos(\alpha) = \frac{1-\sqrt{v}}{\sqrt{u}},
\end{equation} 
they can be expanded in an absolutely convergent series in the Gegenbauer polynomials \cite{Hogervorst2013, Costa2016}:
\begin{equation}
    G_{\mathcal{O}}(\tau, \alpha) = \sum_{n\in 2\mathbb{N}_0} e^{-(\Delta_{\mathcal{O}}+n)\tau} \sum_{j} B_{n, j}(\Delta_{\mathcal{O}}) C_j^{d/2-1}(\cos\alpha),
    \label{eq:Gexpansion}
\end{equation}
where in \(d=3\) dimensions the Gegenbauer polynomials correspond to the Legendre polynomials \(C_j^{d/2-1}(\cos\alpha)=P_j(\cos\alpha)\).
The \(j\) in the second sum is determined by the usual rules for the addition of angular momenta between the spin \(l\) of the primary \(\mathcal{O}\) and \(n\),
\begin{equation}
    j \in \{ \max(0, l-n), \cdots,  l+n\},
    \label{eq:angmomaddition}
\end{equation}
and the \(B_{n, j}\) are normalization coefficients. In three dimensions, they are normalized by 
\begin{equation}
\label{eq:Bnorm}
   B_{0, j}(\Delta_{\mathcal{O}}) = 4^{\Delta_{\mathcal{O}}} \frac{l!}{(1/2)_l} \delta_{j, l},
\end{equation}
where \(()_l\) is the Pochhammer symbol. For their recursive calculation, we used the Mathematica Notebook provided with Ref.\,\cite{Costa2016}.  

\subsection{The antipodal scalar four-point function on \texorpdfstring{\(\mathbb{R}\times S^2\)}{RxS2}}
\label{subsec:antipodal4pt}

\begin{figure}[ht]
    \centering
    \includegraphics[width=0.49\textwidth]{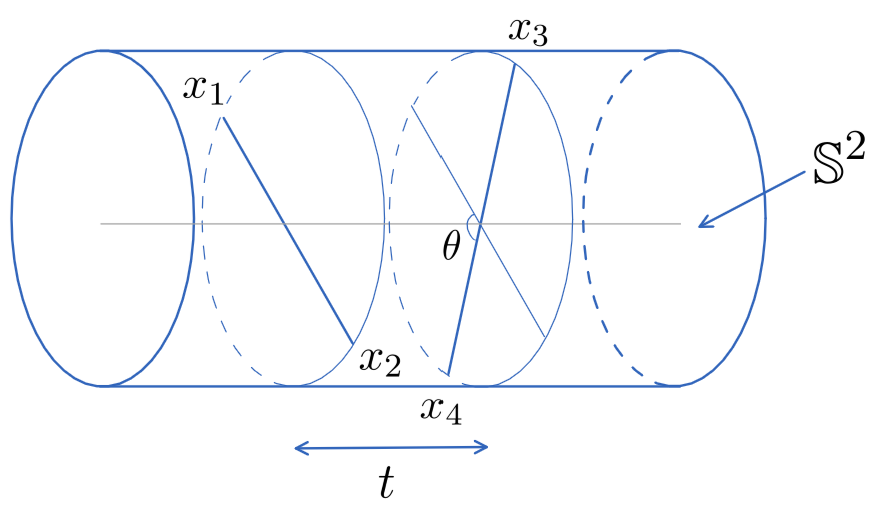}
    \caption{Schematic illustration of the antipodal frame on \(\mathbb{R}\times S^2\) in which we calculate the 4-point function}
    \label{fig:antipodal_frame}
\end{figure}

While in Euclidean space \(\alpha\) and \(\tau\) are not accessible, that changes if we radially quantize the theory.  If we measure the four-point function on \(\mathbb{R}\times S^2\) with \(x_i=(t_i, \vec{n_i}), \ i\in\{1, 2, 3, 4\}\) placed such that they lie pairwise at identical times, \(t_1=t_2\) and \(t_3=t_4\), and on antipodal points of \(S^2\), \(\vec{n}_1=-\vec{n}_2\) and \(\vec{n}_3=-\vec{n}_4\) (see Fig.~\ref{fig:antipodal_frame}), we have \cite{Brower:2020jqj}
\begin{equation}
\vec{n_1}\cdot \vec{n_4} \equiv \cos(\theta) = \cos(\alpha), \ \ \ \cosh(t) = \cosh(\tau)
\end{equation} 
such that we can directly measure
\begin{align}
    g(t, \cos\theta) 
    = 1 + \sum_{\mathcal{O}} f_{\sigma \sigma \mathcal{O}}^2 \sum_{n\in 2\mathbb{N}_0}e^{-(\Delta_{\mathcal{O}}+n)t} \\ \nonumber
    \times \sum_{j} B_{n, j}(\Delta_{\mathcal{O}}) P_j(\cos\theta).
    \label{eq:OPEexpthetat}
\end{align}
This is the big advantage of doing lattice calculations in radial quantization compared to usual Euclidean lattices. Now \(g(t, \cos\theta) \) has the form of a partial wave expansion in the chosen antipodal frame
\begin{equation}
g(t, \cos\theta) = 1 + \sum_{j} c_j(t) P_j(\cos\theta)
\label{eq:partial_wave}
\end{equation}
with expansion coefficients 
\begin{align}
c_j(t) = \sum_{\mathcal{O}} f_{\sigma \sigma \mathcal{O}}^2 \sum_{\substack{n\in2\mathbb{N}_0 \\ n\geq|j-l|}}^{\infty} B_{n,j}(\Delta_{\mathcal{O}}) e^{-(\Delta_{\mathcal{O}}+n)t} 
\label{eq:expansion_coeff}
\end{align}
so that each expansion coefficient consists of infinite towers of exponentials for all even parity and even spin primaries. By fitting these expansion coefficients, we can extract the quantities \(\Delta_{\mathcal{O}}\) and \(f_{\sigma \sigma \mathcal{O}}\).

\section{Radially quantizing the 3d Ising CFT with Quantum Finite Elements}

\begin{figure}[ht]
    \centering
    \includegraphics[width=0.48\textwidth]{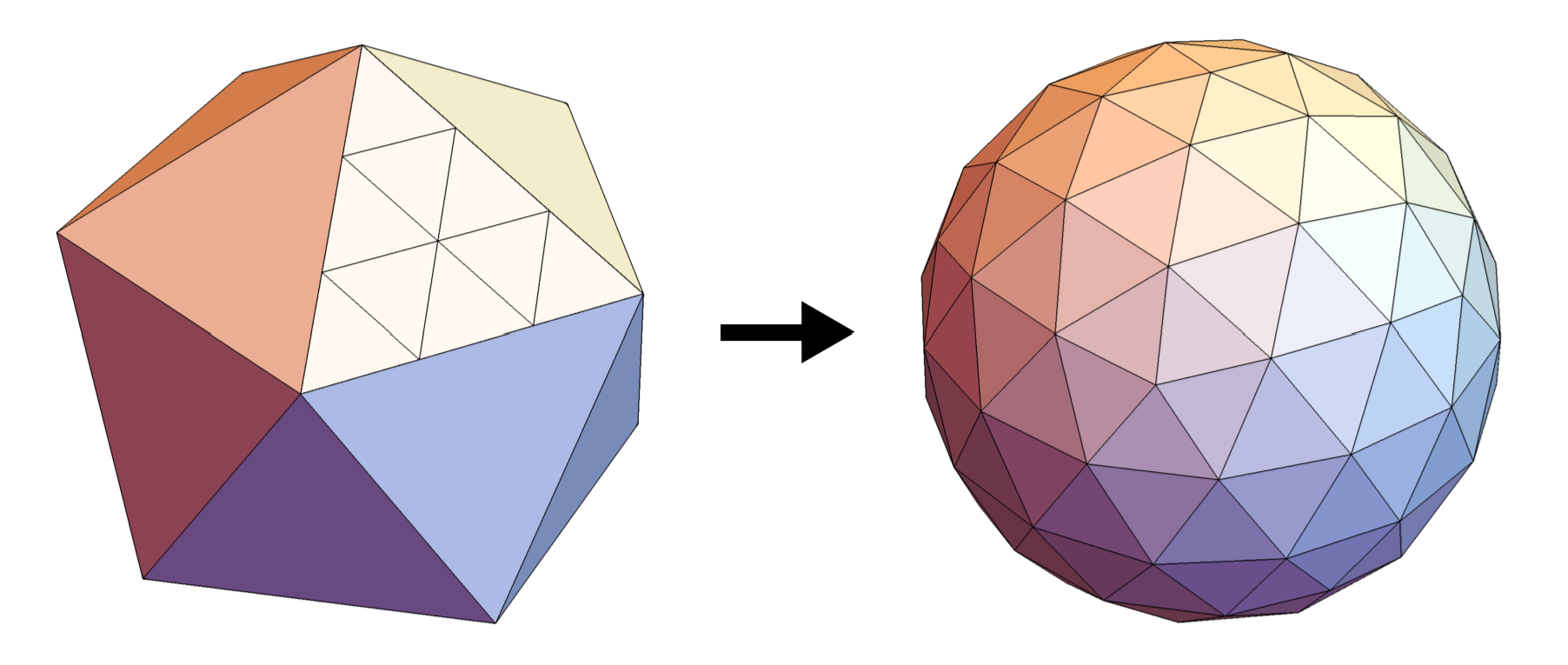}
    \caption{Construction of the simplicial lattice for refinement \(L=3\)}
    \label{fig:simplicial_complex}
\end{figure}

In Refs.~\cite{Brower2018, Brower:2020jqj}, Brower \textit{et al.}\ succeeded at radial lattice quantization of \(\phi^4\) theory. For that, they approximated \(S^2\) via a series of simplicial lattices as shown in Fig.~\ref{fig:simplicial_complex}.

Starting from an icosahedron, its edges are first subdivided into \(L\) equal pieces, thereby introducing a finer triangulation. Subsequently, the \(N=10L^2+2\) vertices are projected onto the sphere. Using the finite element method, the classical action can then be discretized on this simplicial complex \cite{Brower:2020jqj}:
\begin{align}
    S = &\frac{1}{2} \sum_{t, x} \sum_{y\in \langle x, y \rangle} \frac{l^*_{xy}}{l_{xy}} \left(\tilde{\phi}_{t,x} - \tilde{\phi}_{t,y}\right)^2 + \frac{a^2}{4R^2}\sqrt{\tilde{g}_x}\tilde{\phi}_{t,x}^2  \label{eq:action_discrete} \\ 
    &+\sqrt{\tilde{g}_x} \left[ \frac{a^2}{a_t^2} \left(\tilde{\phi}_{t,x} - \tilde{\phi}_{t+1,x}\right)^2 + m_0^2 \tilde{\phi}_{t,x}^2 + \lambda_0 \tilde{\phi}_{t,x}^4 \right], \nonumber
\end{align}
where we sum over \(x \in \{1, 2, \cdots, N\}\) on each sphere and \(t \in \{1, 2, \cdots, L_t\}\) along the cylinder with periodic boundary conditions. Here, \(\tilde{\phi}_{t,x}\) is the dimensionless field, \(m_0^2\) and \(\lambda_0\) the dimensionless mass and coupling. \(a\) and \(a_t\) are the average lattice spacings on the sphere and along the cylinder, respectively, and \(a\) is related to the radius \(R\) of the sphere via 
\begin{equation}
    a^2/R^2 = \frac{8\pi}{\sqrt{3}N} = \frac{8\pi}{\sqrt{3}(10L^2+2)} .
    \label{eq:aR}
\end{equation}
The bare speed of light \(a/a_t\) is set to one, though it is renormalized by interactions. For further details on the construction of this action as well as on how the dimensions of the quantities are restored, see Refs.~\cite{Brower2018, Brower:2020jqj}.

While the position-dependent finite element weights \(l^*_{xy}/l_{xy}\) and \(\sqrt{\tilde{g}_x}\) ensure the convergence of the action in Eq.~\eqref{eq:action_discrete} to the spherically symmetric classical continuum theory \cite{Brower2018}, Brower \textit{et al.}\ found that for the quantum theory, they introduce fluctuations in the effective cutoff that are amplified by UV divergences, preventing convergence to the spherically symmetric quantum continuum theory \cite{Brower:2020jqj}. To compensate for these quantum UV defects, perturbative counterterms, the so-called Quantum Finite Elements, were introduced to the lattice action:
\begin{equation}
    S_{QFE} = S - \sum_{t, x} \sqrt{\tilde{g}_x}\left(6\lambda_0\delta G_{t, x} - 24 \lambda_0^2 \delta G_{t, x}^{(3)}\right) \tilde{\phi}^2_{t, x},
    \label{eq:QFEaction}
\end{equation}
where \(\delta G_{t, x}\) and \(\delta G_{t, x}^{(3)}\) are calculated numerically from the free lattice propagator \(G_{t, x; t', y}\) via
\begin{align}
    \delta G_{t, x} &\equiv G_{t, x; t, x} - \frac{1}{N} \sum_{x'=1}^N \sqrt{\tilde{g}_{x'}} G_{t, x'; t, x'}, \\
    \delta G_{t, x}^{(3)} &\equiv \sum_{t', y} \sqrt{\tilde{g}_y} \left(G_{t, x; t', y}^3 - \frac{1}{N} \sum_{x'=1}^N \sqrt{\tilde{g}_{x'}} G_{t, x'; t', y}^3\right). 
\end{align}

\noindent In order to tune this QFE action to the Ising critical surface, the Binder cumulant was studied for fixed \(\lambda_0=0.2\), finding \(m_0^2=-\mu_0^2=-0.27018(4)\) as the critical coupling. With the parameters tuned to these values, the study of the scalar two-point function carried out in \cite{Brower:2020jqj} suggests that we reach the 3d Ising CFT in the continuum limit \(a/R\rightarrow 0\). In the following, we will fix the bare parameters to these values for our analysis of the interacting theory. 

\section{\label{sec:interacting} Calculating and fitting the antipodal lattice four-point function}

To compute the antipodal scalar four-point function on the lattice, we used our QFE action \eqref{eq:QFEaction} and carried out Monte Carlo simulations with the Brower-Tamayo cluster algorithm \cite{Brower:1989mt} combined with Metropolis \cite{Metropolis:1953am} and overrelaxation \cite{Adler:1981sn, Whitmer:1984he, Brown:1987rra, Creutz:1987xi}. After some initial equilibration sweeps, \(\mathcal{O}\)(volume) four-point function configurations of our scalar lattice fields \(\tilde{\phi}\) were sampled in each sweep along with the corresponding equal-time 2-point functions to obtain \(g(u, v)\) as in Eq.~\eqref{eq:guv}. Subsequently, we projected onto Legendre polynomials to obtain the expansion coefficients \(c_j(t)\) in Eq.~\eqref{eq:Gexpansion}. Note that for finite lattice spacing, the lattice operator \(\tilde{\phi}\) does not exactly correspond to the primary \(\sigma\) as in Eq.~\eqref{eq:Gexpansion} but it is a mixture of all \(\mathbb{Z}\)-even scalar primaries of the CFT. However, as \(a/R\rightarrow 0\), the contribution of operators other than \(\sigma\) go to zero. 

\begin{figure}[ht]
    \centering
    \includegraphics{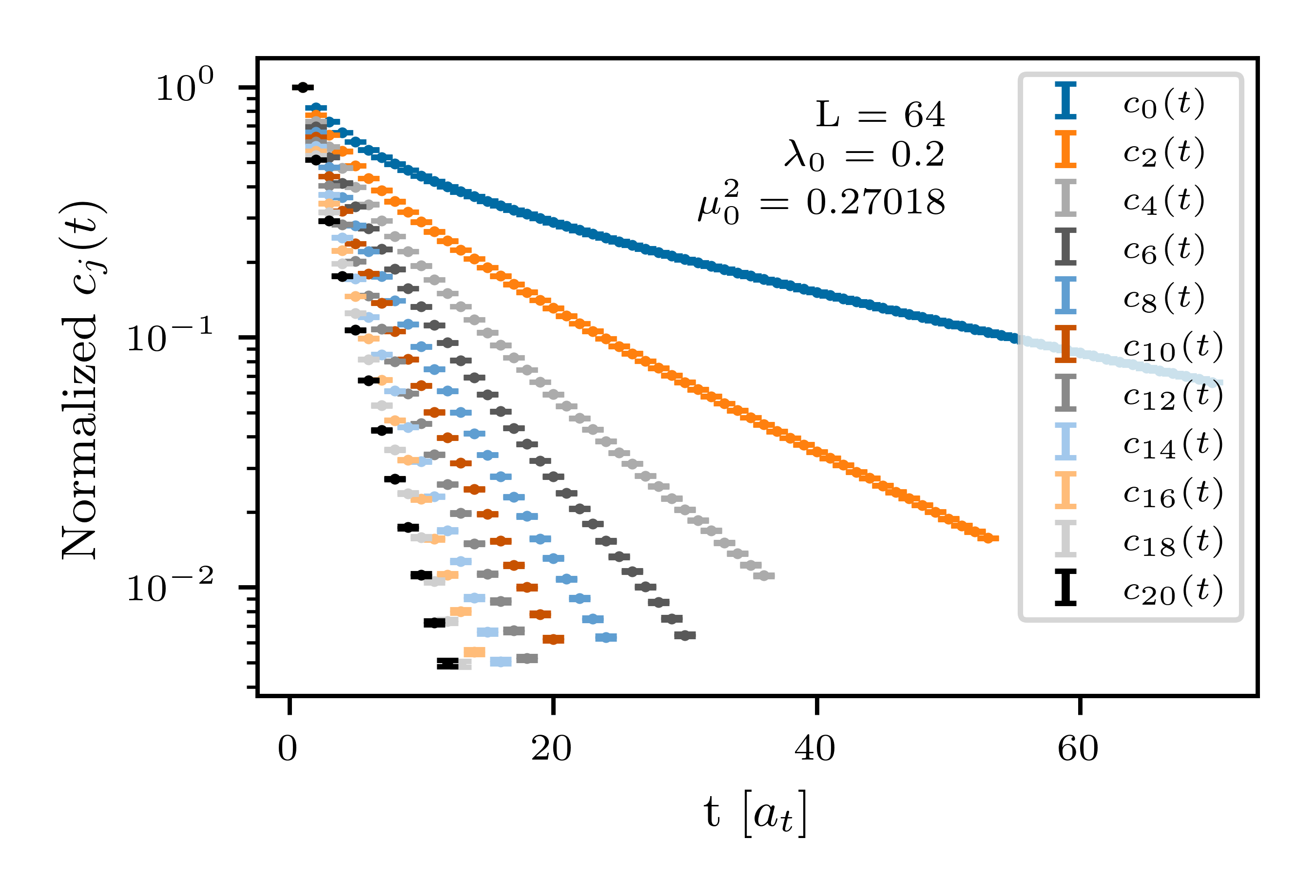}
    \caption{Monte Carlo results for the four-point function expansion coefficients \(c_j(t)\) from \(N=800\) independent runs with statistical errors. The coefficients are plotted for a \(t\)-range such that the relative error of the effective mass does not exceed \(\Delta m_{eff}/m_{eff} \leq 0.125\).}
    \label{fig:cj}
\end{figure}

Such simulations were carried out for different lattice refinements \(L\in\{24, 28, 32, 36, 40, 44, 48, 56, 64\}\) in order to take the continuum limit later by sending \(L\rightarrow \infty\), corresponding to \(a/R\rightarrow 0\). We performed \(N=8000\) independent simulations for \(L=24\), \(N=1600\) for \(32\leq L \leq 56\), and \(N=800\) for \(L=64\). The resulting data is freely available in the Zenodo repository associated with this project \cite{Zenodo}. For \(L=64\), the measured expansion coefficients up to \(j=20\) are shown in Fig.~\ref{fig:cj} with the error bars determined by the statistical error from averaging over the \(N\) independent runs. This figure illustrates that even for the \(L\)-value with the least amount of statistics, we have several timeslices of good data up until high \(j\), such that it should in principle be possible to extract information about high-spin operators. 

And in fact, by simply fitting unconstrained sets of exponentials to the different \(c_j\), we could extract five exponentials for \(j=0\), and even for \(j=10\) we could fit up to 3 exponentials. However, the statistics did not allow, especially for the higher \(j\), to always resolve the contribution from different primaries. For example, in the case of \(j=2\), we expect from Eq.~\eqref{eq:expansion_coeff} and the bootstrap results in Tab.~\ref{tab:bsvalues} that the leading exponential \(\propto e^{-\Delta_{T}t}\) involves the scaling dimension \(\Delta_T\) of the energy-momentum tensor \(T\) with \(\ell=2\), while the next-lowest exponential \(\propto e^{-(\Delta_{\epsilon}+2)t}\) should stem from the \(\ell=0\) operator \(\epsilon\). To be able to resolve these two contributions at \(L=64\) without having to introduce additional constraints in the fitting parameters such as Eq.~(\ref{eq:expansion_coeff}), we would likely have to increase the statistics by a factor 9-16, which is computationally not feasible as already the \(N=800\) runs took \(\mathcal{O}(6\times10^5)\) hours of computing time on single cores on the BU shared computing cluster. 

\begin{table}[]
    \centering
    \begin{tabular}{cll}
    \hline \hline
         Operator \(\mathcal{O}\) (\(l^P\)) & \(\Delta_{\mathcal{O}}\) & \(f_{\sigma \sigma \mathcal{O}}\) \\
         \hline
         \(\epsilon\) (\(0^+\)) & 1.41265(36) & 1.05185(12)\\
         \(\epsilon'\) (\(0^+\)) & 3.82951(61) & 0.05304(16)\\
         \(T\) (\(2^+\)) & 3 & 0.32613776(45) \\
         \(T'\) (\(2^+\)) & 5.499(17) & 0.01054(10) \\ 
    \hline \hline
    \end{tabular}
    \caption{Conformal bootstrap values for scaling dimensions and OPE coefficients of conformal primaries \(\epsilon, T, \epsilon'\), and \(T\) \cite{Reehorst2022, SimmonsDuffin2016}} 
    \label{tab:bsvalues}
\end{table}

Thus, instead of fitting each \(c_j\) independently with unconstrained exponentials, we analyzed the Monte Carlo data by fitting the \(c_j\) simultaneously for different \(j\) with much more constrained functions based on the expansion Eq.~\eqref{eq:expansion_coeff}. This has the advantage that adding a new primary operator \(\mathcal{O}\) to the fit only introduces two new fit parameters in total, namely \(f_{\sigma \sigma \mathcal{O}}\) and \(\Delta_{\mathcal{O}}\), while describing (in principle infinite) towers of exponentials for each \(c_j\). The scaling dimension and OPE coefficients of a spin-\(l\) operator are then mainly determined by the operator's contribution to \(c_{\ell}\) while appropriately contributing to higher-order terms in \(c_j\) with \(l \neq j\). Overall, this procedure improves convergence and helps resolve the different contributions. 

Of course, strictly speaking, Eq.~\eqref{eq:expansion_coeff} is only true in the continuum limit as our finite lattice spacing breaks conformal invariance. However, we still performed fits based on the form of Eq.~\eqref{eq:expansion_coeff} because we expect that if \(a/R\) is small enough, the corrections due to the breaking of lattice spacing can be absorbed into corrections to our fit parameters that vanish as \(a/R\rightarrow 0\). 

In the present proof of principle, we focused on simultaneous fits to \(c_0\) and \(c_2\), for which the leading contributions come from the primaries \(\epsilon\) \((0^+)\) and the energy-momentum tensor \(T\) \((2^+)\), respectively. Furthermore, we included the first subleading operators for each \(\ell\), namely \(\epsilon'\) \((0^+)\) and  \(T'\) \((2^+)\) in our fits. We truncated the sums of exponentials at \(n_{max}=20\) for \(\epsilon\) and correspondingly lower values for the other operators such that the largest exponents are of the same order of magnitude. With these truncations, Eq.~\eqref{eq:expansion_coeff} yields the following fit functions for \(c_0(t)\) and \(c_2(t)\) (in terms of the lattice ``time'' \(t\))

\begin{align}
    \label{eq:c0_fit}
    c_0^{fit}(t) =  &\sum_{n=0}^{n_{max}} f_{\sigma \sigma \epsilon}^2 B_{n, 0}(\Delta_{\epsilon}) e^{-(\Delta_{\epsilon}+n)ta_t/R} \\ \nonumber
    & + \sum_{n=0}^{n_{max}-2} f_{\sigma \sigma \epsilon'}^2 B_{n, 0}(\Delta_{\epsilon'}) e^{-(\Delta_{\epsilon'}+n)ta_t/R} \\ \nonumber
    & + \sum_{n=2}^{n_{max}-4} f_{\sigma \sigma T'}^2 B_{n, 0}(\Delta_{T'}) e^{-(\Delta_{T'}+n)ta_t/R}\\ \nonumber
    &+ (t\rightarrow L_t-t),
\end{align}
\begin{align} 
    \label{eq:c2_fit}
    c_2^{fit}(t) = & \sum_{n=0}^{n_{max}-2} f_{\sigma \sigma T}^2 B_{n, 2}(\Delta_{T}) e^{-(\Delta_{T}+n)ta_t/R} \\ \nonumber
     + &\sum_{n=0}^{n_{max}-4} f_{\sigma \sigma T'}^2 B_{n, 2}(\Delta_{T'}) e^{-(\Delta_{T'}+n)ta_t/R}\\ \nonumber
     + &\sum_{n=2}^{n_{max}} f_{\sigma \sigma \epsilon}^2 B_{n, 2}(\Delta_{\epsilon}) e^{-(\Delta_{\epsilon}+n)ta_t/R} \\ \nonumber
     + &\sum_{n=2}^{n_{max}-2} f_{\sigma \sigma \epsilon'}^2 B_{n, 2}(\Delta_{\epsilon'}) e^{-(\Delta_{\epsilon'}+n)ta_t/R} \\ \nonumber
     &+ (t\rightarrow L_t-t), 
\end{align}
where the quantities \(\Delta_{\mathcal{O}}\) and \(f^2_{\sigma \sigma \mathcal{O}}\) are taken as the fit parameters. Note that we did not include contributions from the energy-momentum tensor \(T\) to \(c_0(t)\) because for the theoretical continuum value \(\Delta_T = 3\), all \(B_{n, 0}(3)\) vanish. In order to convert the lattice time of our data, which is given in units of \(a_t\), to the time on \(\mathbb{R}\times S^2\) in Eq.~\eqref{eq:expansion_coeff}, which after the Weyl transform has units of \(R\), we included factors \(a_t/R\) in the exponentials. This can be written as \(a_t/R = a/(Rc_R)\) where \(a/R\) is known from Eq.~\eqref{eq:aR} and \(c_R\) is the renormalized speed of light \(c_R = a/a_t\). \(c_R\) was determined non-perturbatively in Ref.~\cite{Brower:2020jqj}, yielding \(c_R = 0.996\). Furthermore, to take into account wraparound the effects due to the periodicity along the cylinder to leading order, we added for each exponential in our fit function the same term with \(t\) substituted by \(L_t-t\).

\begin{figure}[ht]
    \centering
\includegraphics{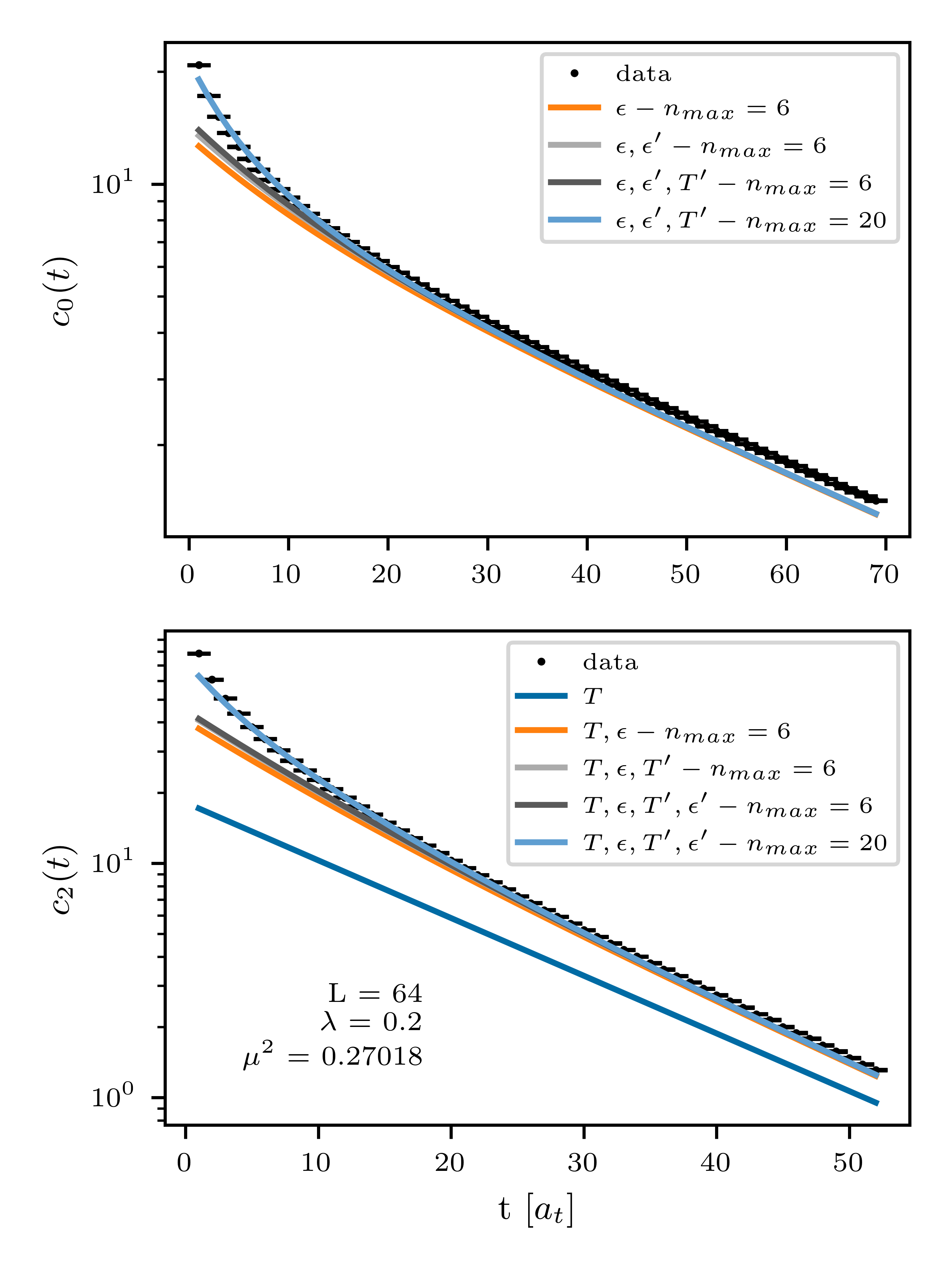}
    \caption{Lattice data for $c_0$ and $c_2$ at a refinement of \(L=64\) compared to theoretical curves based on truncations of Eq.~\eqref{eq:expansion_coeff}. For each curve, we include another primary operator contribution, corresponding to a new line in Eqs.~\eqref{eq:c0_fit} and \eqref{eq:c0_fit}, or a higher \(n_{max}\). The light blue curve corresponds to the function we used for the fits in our analysis, including operators \(\epsilon, \epsilon', T\), and \(T'\) with \(n_{max}=20\). Note that these are not fits to the data but theoretical curves with the bootstrap values as scaling dimensions and OPE coefficients. }
    \label{fig:fit_funcs}
\end{figure}

The light blue curves in Fig.~\ref{fig:fit_funcs} show the functions Eqs.~\eqref{eq:c0_fit} and \eqref{eq:c2_fit} plotted with the bootstrap values for \(\Delta_{\mathcal{O}}\) and \(f_{\sigma \sigma \mathcal{O}}\), alongside our lattice data for \(L=64\). Comparison of these theoretical continuum curves with our lattice data shows apparent consistency, justifying our use of fit functions of this form. The other curves illustrate how the fit functions change depending on how many operators and \(n\) are included. Evidently, the operators \(\epsilon\) and \(T\) are most influential while including the subleading \(\epsilon'\) and \(T'\) only slightly shift the theoretical curve, mostly at low \(t\)-values. However, including these operators in fits decreased excited state contamination of the fit parameters for the leading operators and significantly increased the model probability of our fits, which we determined using the Akaike Information Criterion as described in Ref.~\cite{Jay2020}. Including even higher order operators like \(\epsilon''\) led to either unconstrained or unphysical fit parameters, however. Thus, for simultaneous fits of \(c_0\) and \(c_2\), we determined the fits including \(\epsilon, \epsilon', T\) and \(T'\) (Eqs.~\eqref{eq:c0_fit} and \eqref{eq:c2_fit}) to be optimal.

Because of the truncations and approximations made in our fit functions, we do not expect them to describe our data within the entire \(t\)-range. Thus, we performed a model averaging procedure as described in \cite{Jay2020} in order to determine a reasonable range of minimal timeslices \((t_0^{min}, t_2^{min})\). For each value of \(L\), the fitting procedure can then be summarized as follows:

\begin{itemize}
    \item First, we performed simultaneous fits to \(c_0(t)\) and \(c_2(t)\) with the respective fit functions Eq.~\eqref{eq:c0_fit} and Eq.~\eqref{eq:c2_fit} with fixed maximal timeslices \((t_0^{max}, t_2^{max})\) and all possible combinations of starting times \((t_0^{min}, t_2^{min})\). Fig.~\ref{fig:fit_funcs} suggested the use of the bootstrap values listed in Tab.~\ref{tab:bsvalues} as initial guesses for our fit parameters to speed up the convergence of the non-linear fitting procedure. However, we ensured that there is no dependence of our results on the initial guesses by repeating the fits with the first fit results as initial guesses and with checks perturbing these initial guesses. \((t_0^{max}, t_2^{max})\) were chosen such that the relative error of the effective mass for \(c_0\) and \(c_2\) never exceeded \(12.5\%\). The simultaneous fits were carried out with the least squares method using the L-BFGS-B optimization algorithm implemented in SciPy~\cite{Byrd1995}. 

    \item Then, we calculated the model probabilities for all fits with the Akaike criterion according to \cite{Jay2020}.
    
    \item Thereafter, we discarded fits with unphysical~\footnote{For \(L=36\) some of the most probable fits had unphysical values of \(\Delta_{\epsilon'} < \Delta_T\), which we discarded.} or unconstrained parameters (i.e. parameters with errors higher than \(50\%\)) as well as fits with \((t_0^{min}, t_2^{min})\) far from the values with highest model probability. After eliminating these fits, we renormalized the model probability.
    
    \item For each fit parameter, we lastly computed a weighted average over all fits with renormalized model probability higher than $0.1\%$.

The results of the individual fits used in calculating the final model averaged values for each \(L\) are available in the Zenodo repository associated with this project \cite{Zenodo}.

\end{itemize}

\section{\label{sec:results}Results}

\begin{figure*}[ht]
\includegraphics{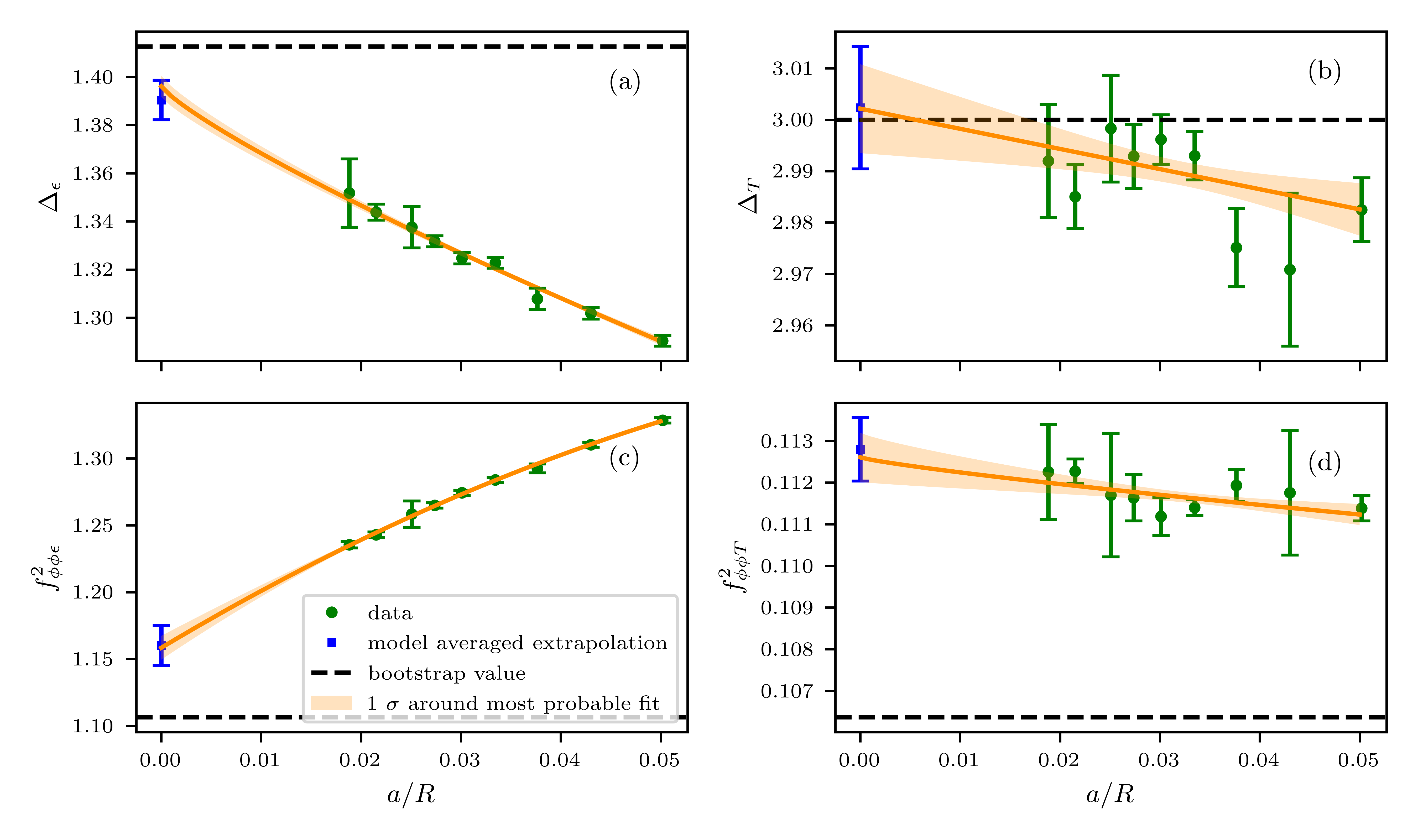}
\caption{Model averaged fit results (green dots) for the scaling dimensions and OPE coefficients of the leading operators \(\epsilon\) and \(T\) as a function of the lattice spacing. We have performed extrapolations to \(a/R\rightarrow0\)  of different functional forms \eqref{eq:FSS}-\eqref{eq:quadratic} and including different ranges of \(a/R\). The values shown as blue squares at \(a/R=0\) correspond to the continuum results obtained after model averaging all different extrapolation fits. For each quantity, we also show the extrapolation fit with the highest model probability in orange - in (a) and (d), this is the finite-size scaling based fit, for (b) it is the linear extrapolation, and for (c) the quadratic fit, always including all data points. For comparison, the continuum bootstrap values are shown as dashed black lines.}
\label{fig:leading}
\end{figure*}

\begin{table*}[]
    \centering
    \begin{tabular}{cccccc}
    \hline\hline
         Quantity & Bootstrap & FSS Fit & Fit with highest \(p_{model}\) & Fit Model Average & Deviation \\
         \hline
         \(\Delta_{\epsilon}\) & 1.41265(36)
         & 1.3961(46) & 1.3961(46) & 1.3905(83) & 2.7\,\(\sigma\) \\
          \(f_{\sigma \sigma \epsilon}^2\) & 1.10639(26) 
          & 1.1625(31) & 1.1583(91) & 1.160(15) & 3.6\,\(\sigma\) \\
         \(\Delta_{T}\) & 3 (exact) & 3.005(11) & 3.0021(86) & 3.002(12) & 0.2\,\(\sigma\) \\
         \(f_{\sigma \sigma T}^2\) & 0.10636583(30)
         & 0.11261(58) & 0.11261(58) & 0.11280(76) & 8.5\,\(\sigma\)\\ 
         \(\Delta_{\epsilon'}\) & 3.82951(61) 
         & 3.82(24) & 3.74(20) & 3.78(27) & 0.2\,\(\sigma\) \\
         \(f_{\sigma \sigma \epsilon'}^2\) & 0.0028102(59) & 0.0020(24)
          & 0.0027(20) & 0.0028(29) & 0.02\,\(\sigma\) \\
         \(C_T/C_T^{(free)}\) & 0.946543(42)
         & 0.896(10) & 0.8961(81) & 0.896(11) & 3.9\,\(\sigma\)\\
         \(f_{\sigma \sigma \epsilon}^2/f_{\sigma \sigma T}^2\) & 10.4017(24) & 10.312(79) & 10.19(12) & 10.30(16) & 0.6\,\(\sigma\) \\
         \(C_T/C_T^{(free)} \cdot f_{\sigma\sigma\epsilon}^2/\Delta_{\epsilon}^2\) & 0.52478(30) 
         & 0.5219(81) & 0.5219(81) & 0.528(16) & 0.2\,\(\sigma\)\\
         \hline\hline
    \end{tabular}
    \caption{Extrapolation results. We show the continuum values obtained with the finite-size scaling based fits including all data points, the results from the fits with the highest model probability, and the model averaged extrapolations for all quantities that are also plotted in Figs.~\ref{fig:leading}-\ref{fig:C_T}. These values are compared to those obtained from the conformal bootstrap \cite{Reehorst2022, SimmonsDuffin2016}, and in the last column, we show the deviation of the model averaged extrapolation values to the bootstrap results with respect to the statistical and fitting errors of our lattice results. }
    \label{tab:extra_res}
\end{table*}

The model averaged fit results for the scaling dimensions and OPE coefficients of the leading operators are shown in Fig.~\ref{fig:leading} as functions of the dimensionless lattice spacing \(a/R\). The errors were calculated from the covariances of the fits according to Eq.~(17) in \cite{Jay2020}. Thus, they do not capture possible systematic errors in the setup of the QFE framework. 

If we compare our lattice results to the values obtained by the conformal bootstrap (shown in Tab.~\ref{tab:bsvalues} and plotted as black dotted lines), it is apparent that \(\Delta_{\epsilon}\), \(f_{\sigma\sigma\epsilon}^2\), and \(\Delta_{T}\) trend towards the continuum bootstrap values as \(a/R\rightarrow 0\). Only \(f_{\sigma \sigma T}^2\), the OPE coefficient associated with the energy-momentum tensor, does not show such a clear trend.

For a quantitive comparison, we had to extrapolate our lattice results to the continuum limit \(a/R\rightarrow 0\). In Appendix~\ref{sec:fss}, we performed a finite-size scaling analysis of \(g(u, v)\), yielding that to leading order, the squared OPE coefficients and scaling dimensions approach their continuum values with the leading power of \((a/R)^{\Delta_{\epsilon'}-3}\approx(a/R)^{0.83}\). To extrapolate our data to the continuum, we therefore fit our lattice results with functions of this form. Moreover, in the spirit of Taylor series approximations to the true scaling form, we performed linear fits as well as fits including a linear and a quadratic term. Thus, the different fit functions we used for the continuum extrapolations are
\begin{align}
    f_{\text{FSS}} &= c_1(\frac{a}{R})^{0.83} + c_0, \label{eq:FSS} \\
    f_{\text{linear}} &= c_1(\frac{a}{R}) + c_0, \label{eq:linear} \\
    f_{\text{quadratic}} &= c_2 (\frac{a}{R})^{2} + c_1(\frac{a}{R}) + c_0.
    \label{eq:quadratic}
\end{align}
Initially, we had also considered fits to general power laws, \(f_{\text{power}} = c_1(\frac{a}{R})^{c_2} + c_0 \). However, our statistics were not good enough to constrain all fit parameters. Similar problems arose for the quadratic fit, as the only quantity for which such fits could be constrained was \(f_{\sigma\sigma\epsilon}^2\) due to its small error bars. Thus, \(f_{\sigma\sigma\epsilon}^2\) is the only quantity for which we included such quadratic fits in determining our extrapolation results.

To account for the fact that for high \(a/R\) (low \(L\)) we might not be close enough to the continuum for our OPE-based fit functions to be applicable, we also carried out all extrapolating fits excluding the first one or two data points taken at the lowest lattice refinements (\(L=24\) and \(L=28\)). Subsequently, we used model averaging again to weigh the extrapolated values of \(\Delta_{\epsilon}\), \(\Delta_{T}\), \(f_{\sigma \sigma \epsilon}^2\) and \(f_{\sigma \sigma T}^2\) for the different fits by their model probability. Performing this procedure, we saw that in general, the model probability decreased the more low-\(L\) data points we excluded. Moreover, we note that the \(f_{\text{FSS}}\) fit to all data points was among the two extrapolations with the highest model probability for all quantities. 

In Fig.~\ref{fig:leading}, the most probable extrapolation fit (orange) as well as the model averaged continuum extrapolation values (blue) are shown for the scaling dimensions and OPE coefficients of the leading operators. It is apparent that the relative errors of our lattice results are significantly higher for the quantities associated with the \(\ell=2\) operator \(T\) compared to the leading \(\ell=0\) operator \(\epsilon\) due to fact that we have better statistics for \(c_0\), in which \(T\) does not appear, than for \(c_2\). With these relatively high error bars, all of our extrapolations for \(\Delta_T\) (and, thus, in particular the model averaged extrapolation) are in good agreement with the exact value \(\Delta_T=3\). For \(\Delta_{\epsilon}\), the model averaged extrapolation is also just in agreement with the bootstrap value with a deviation of \(2.7\sigma\), even with just taking into account statistical and fitting errors. 

\begin{figure}[ht]
\includegraphics{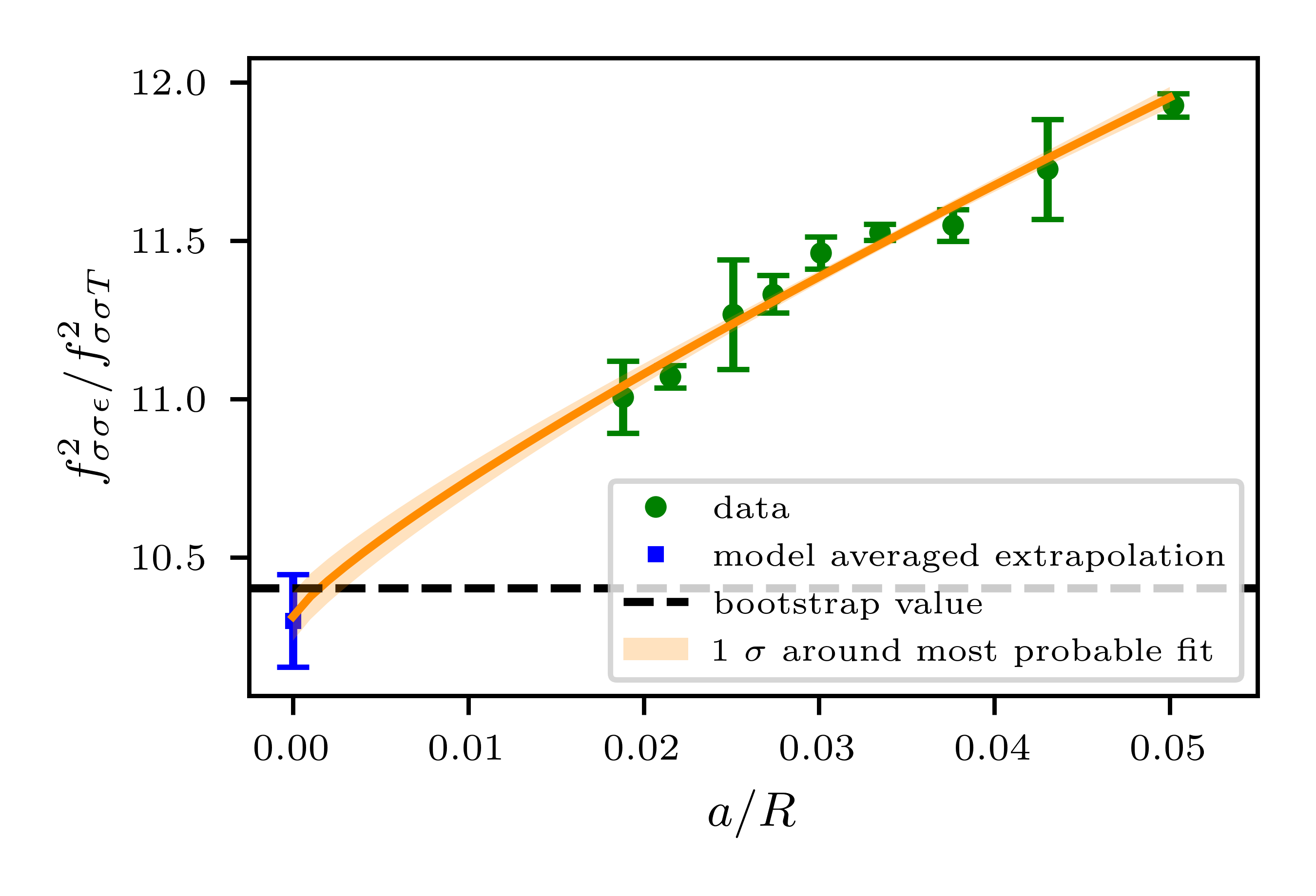}
\caption{Ratio of the model averaged fit results for the OPE coefficients of the leading operators \(\epsilon\) and \(T\) (green dots) as a function of the lattice spacing. We have performed extrapolations to \(a/R\rightarrow0\) according to \eqref{eq:FSS} and \eqref{eq:linear} over different \(a/R\)-ranges. The extrapolation with the highest model probability is the finite-size scaling based fit using the lattice data for \(L\geq 28\), shown in orange. The blue square at \(a/R=0\) represents the continuum value obtained after model averaging all extrapolations. For comparison, the continuum bootstrap value is shown as a dashed black line.}
\label{fig:fr}
\end{figure}

Our continuum extrapolations of \(f_{\sigma \sigma \epsilon}^2\) and \(f_{\sigma \sigma T}^2\), however, are significantly higher than the respective bootstrap values. Given that the value of the OPE coefficients could be sensitive to our normalization choice Eq.~(\ref{eq:Bnorm}) for the conformal blocks, we can explore the sensitivity to this potential source of systematic error
by studying the ratio of the OPE coefficients \(f_{\sigma \sigma \epsilon}^2\) and \(f_{\sigma \sigma T}^2\) in Fig.~\ref{fig:fr}, for which we performed the same extrapolation procedure as for the other quantities. If we take this ratio, all of our extrapolations are in \(1\,\sigma\) agreement with the bootstrap value. This suggests that the normalization of conformal blocks at finite lattice discretization \(L\) warrants further study.

\begin{figure*}[ht]
\includegraphics{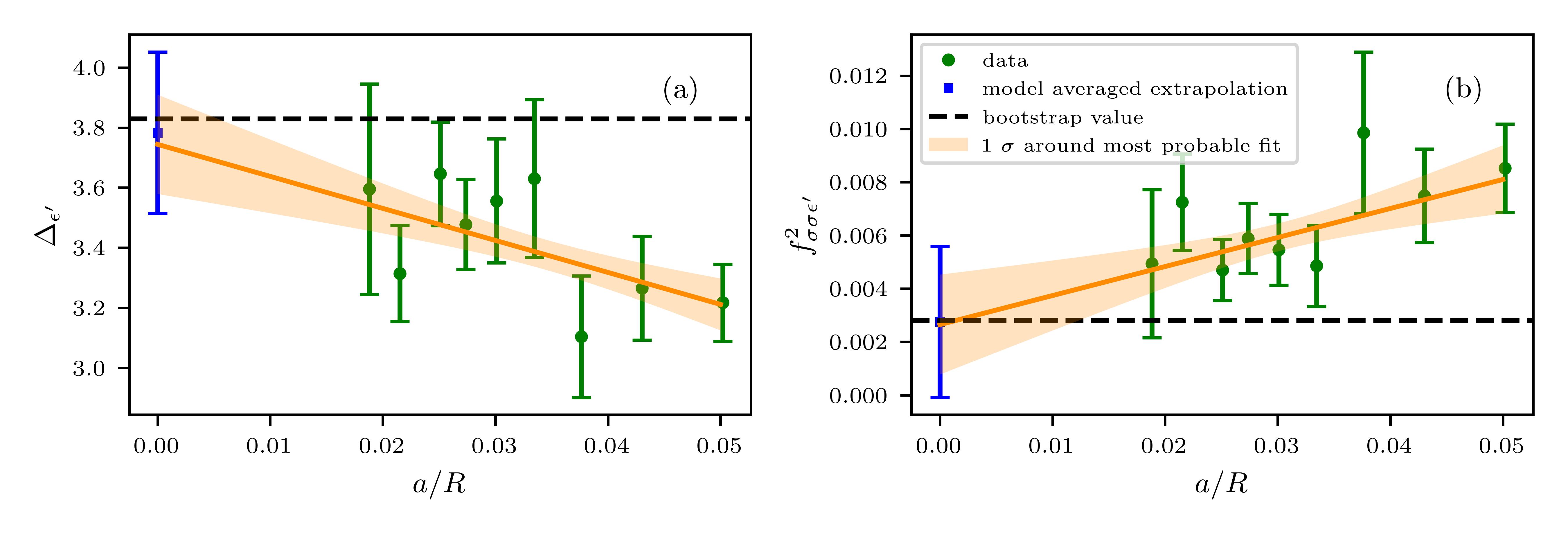}
\caption{Model averaged fit results for the scaling dimension and OPE coefficient of the subleading \(\ell=0\) operator \(\epsilon'\) (green dots) as a function of the lattice spacing. We have performed extrapolations to \(a/R\rightarrow0\) according to \eqref{eq:FSS} and \eqref{eq:linear} over different \(a/R\)-ranges. The extrapolations with the highest model probability are linear fits using the lattice data for all \(a/R\), shown in orange. The blue squares at \(a/R=0\) represent the continuum values obtained after model averaging all extrapolations. For comparison, the continuum bootstrap values are shown as dashed black lines.}
\label{fig:Delta_ep}
\end{figure*}

So far, we have only focused on the fit results for the quantities associated with the leading operators \(\epsilon\) and \(T\) as these have the most signal and the least excited state contamination. To demonstrate that our method also works for the subleading operators, Fig.~\ref{fig:Delta_ep} shows our results for \(\Delta_{\epsilon'}\) and \(f_{\sigma \sigma \epsilon'}^2\), along with extrapolations to \(a/R\rightarrow 0\). The error bars for those quantities are even larger than for \(\Delta_T\) but the extrapolations show agreement with the bootstrap values. Here, we also do not have the issue of a deviation of the OPE coefficient from the bootstrap value as the extrapolated value for \(f_{\sigma \sigma \epsilon'}^2\) has an error of 100\%. To get more precise and reliable results for quantities associated with \(\epsilon'\) and \(T'\), we would have to improve the statistics and go to even higher \(L\) in order to be able to include even more operators in our fits to reduce the excited state contamination of these quantities. 

\begin{figure*}[ht]
\includegraphics{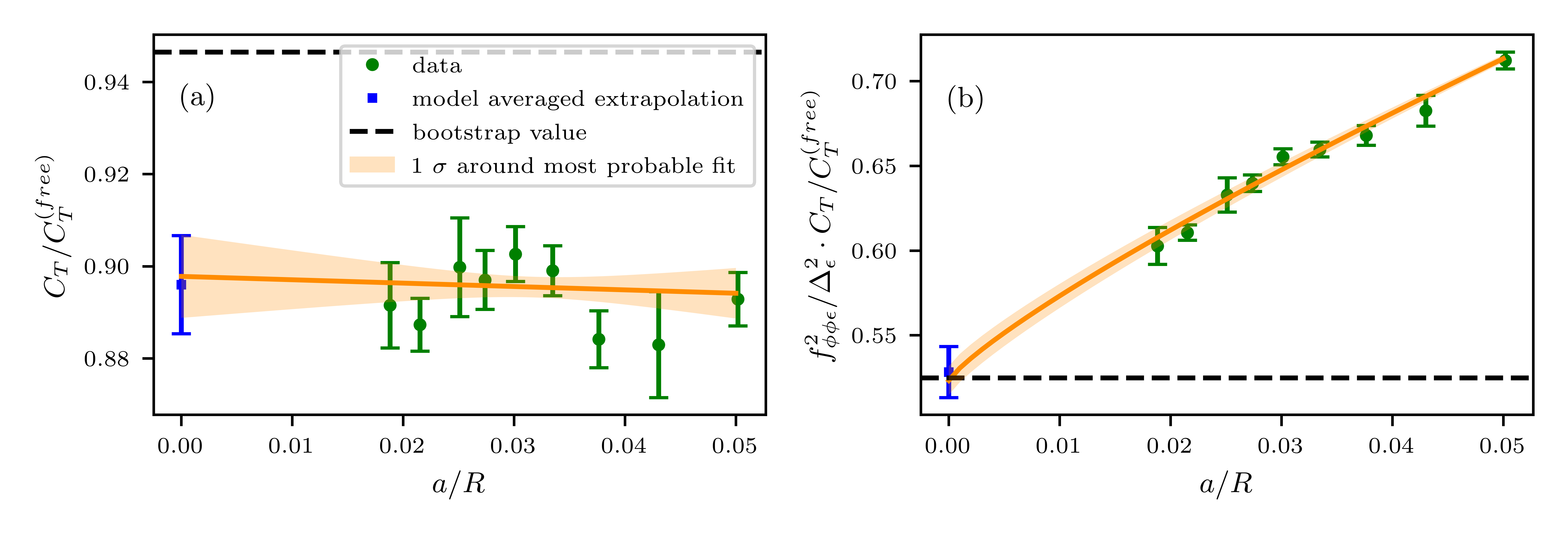}
\caption{(a) Central charge (green dots) in relation to \(C_T^{(free)}=3/2\) as a function of lattice spacing calculated with our the model averaged fit results for \(f_{\sigma \sigma T}^2\) and \(\Delta_T\), as well as \(\Delta_{\sigma} =  0.518(2)\) as obtained in Ref.~\cite{Brower:2020jqj}. (b) \(C_T/C_T^{(free)}\) multiplied by our model averaged fit results results for \(f_{\sigma \sigma \epsilon}^2/\Delta_{\epsilon}^2\) as a function of the lattice spacing. For both quantities, we have performed extrapolations to \(a/R\rightarrow 0\) according to \eqref{eq:FSS} and \eqref{eq:linear} over different \(a/R\)-ranges. The extrapolations with the highest model probability (orange) are a linear extrapolation for \(C_T/C_T^{(free)}\) and a finite-size scaling based extrapolation for \(C_T/C_T^{(free)} \cdot f_{\sigma\sigma\epsilon}^2/\Delta_{\epsilon}^2\), in both cases using all datapoints. The blue squares at \(a/R=0\) represent the continuum values obtained after model averaging all extrapolations. For comparison, the continuum bootstrap values are shown as dashed black lines.}
\label{fig:C_T}
\end{figure*}

Lastly, we want to use our lattice results to determine the central charge of the 3d Ising CFT, which can be calculated via 
\begin{equation}
      C_T = \frac{\Delta_{\sigma}^2\Delta_T^2}{16 f_{\sigma \sigma T}^2}.
     \label{eq:centralcharge}
\end{equation} 
Our results for the central charge relative to the free value \(C_T^{(free)}=3/2\) can be seen in Fig.~\ref{fig:C_T}(a), where we used our lattice values for \(\Delta_T\) and \(f_{\sigma \sigma T}^2\) to calculate \(C_T\) at different \(L\). For \(\Delta_{\sigma}\) we used our continuum value obtained in Ref.~\cite{Brower:2020jqj}. Because our continuum extrapolation for \(f_{\sigma \sigma T}^2\) is significantly too high, it is not surprising that the extrapolation procedure performed for the central charge yields a continuum value that is significantly lower than the bootstrap value. 

If we, however, take an appropriate ratio again, multiplying \(C_T/C_T^{(free)}\propto \Delta_{T}^2/f_{\sigma \sigma T}^2\) by \(f_{\sigma \sigma \epsilon}^2/\Delta_{\epsilon}^2\), the extrapolation yields a continuum value in good agreement with the bootstrap (see Fig.~\ref{fig:C_T}(b)). 

\section{Free CFT analysis of lattice errors}
Errors in the Monte Carlo simulations come from several sources: The statistical Monte Carlo sampling, possible systematic errors with a non-universal form of the action, and the finite lattice errors that need to be removed by extrapolating to infinite lattice volume in the IR and zero lattice spacing for the UV cut-off. 

Clearly, the finite lattize errors can easily be evaluated for the exact free scalar CFT, for which we can calculate the four-point amplitude on the same lattices as in our current $\phi^4$ simulations for the interacting theory. The comparison may be especially revealing here since the 3d Ising CFT is in some ways a small perturbation of the free CFT exploited in the epsilon expansion ($\epsilon = 4 - d$).

For the free theory, the  exact continuum amplitude of the four-point function as in Eq.~\ref{eq:guv}
\begin{align}
g^{(free)}(u,v)= 1 &+ \frac{2}{(2\cosh(t) - 2\cos(\theta))^{\Delta_\sigma^{(free)}}} \nonumber \\
&+ \frac{2}{(2\cosh(t) + 2\cos(\theta))^{\Delta_\sigma^{(free)}}}
\label{eq:4ptamplitudefree}
\end{align}
is a sum over pairs of two-point functions expressed in our radial coordinates 
\begin{equation}
    G^{(2)}_{\phi}(t, \cos \theta) = \dfrac{2}{\left( 2 \cosh t - 2 \cos \theta \right)^{\Delta_{\sigma}^{(free)}}},
\end{equation}
where we now mean the primary operator \(\sigma\) of the free CFT for which \(\Delta_{\sigma}^{(free)}=1/2\).
The partial wave expansion Eq.~\eqref{eq:partial_wave} for the free theory then has the expansion coefficients 
\begin{equation}
    c^{(free)}_{j}(t) = 8 (2 j + 1) \sum_{n=0}^{\infty} e^{-(j+2n+1)t} \dfrac{2^{j}(j+n)!(2 n-1)!!}{n!(2 j + 2 n+1)!!}.
    \label{eq:cfree}
\end{equation}
for the even \(j\) coefficients, while the odd \(j\) coefficients are zero. Moreover, as described in Appendix \ref{app:free}, these equations may be
modified to include the finite length (or finite temperature) of the cylinder. 

We computed the four-point amplitude Eq.~\eqref{eq:4ptamplitudefree} on our lattices,
replacing the free continuum propagator by the finite temperature lattice propagator, calculated by inverting the quadratic action numerically. The goal was then to analyze this synthetic free lattice data similarly to what was done for the interacting case, in order to study how well our lattices enable extrapolations to reproduce the exact free zero-temperature continuum OPE data in Eq.~\eqref{eq:cfree}.
\begin{figure*}[ht]
    \centering
    \includegraphics[width=\textwidth]{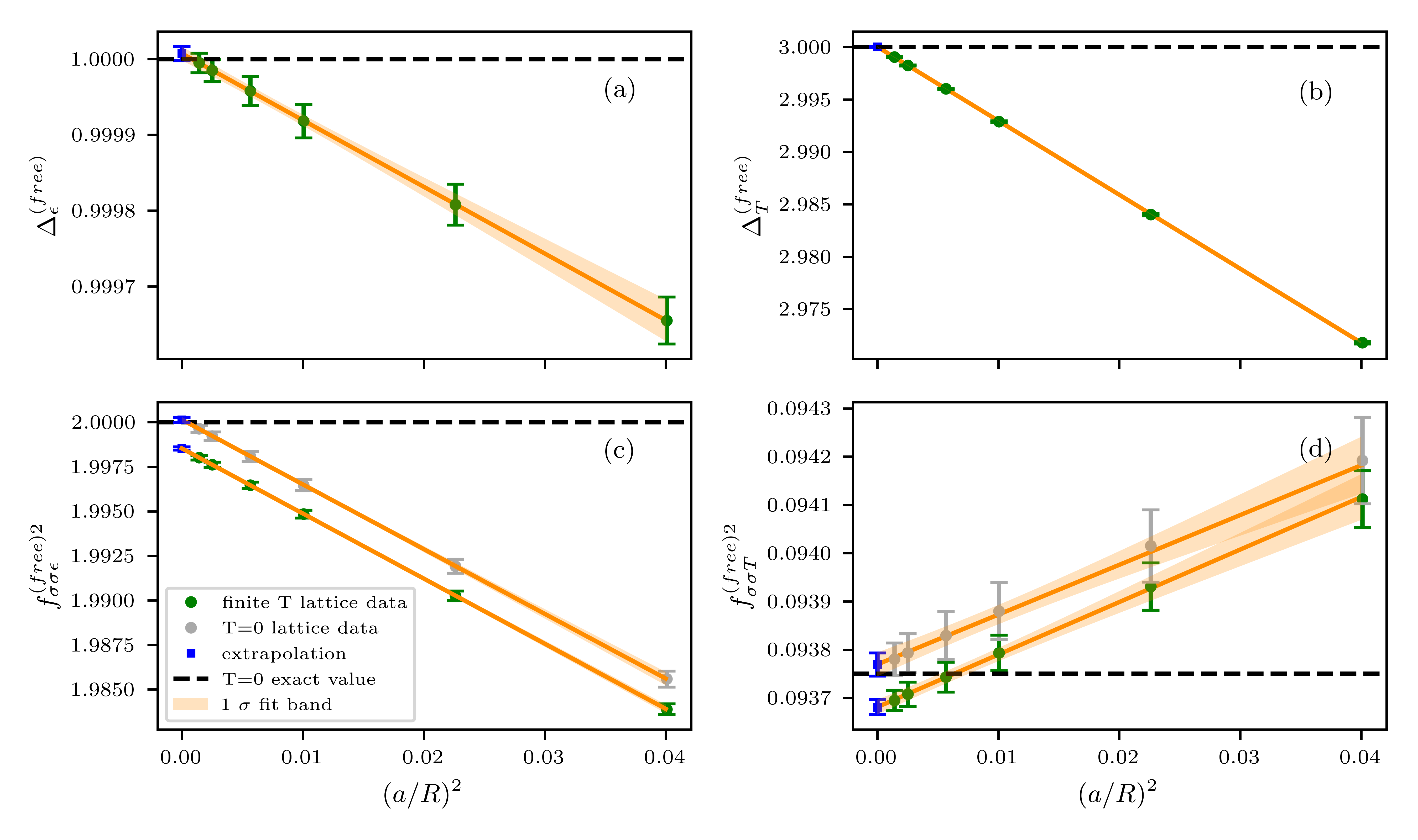}
    \caption{Model averaged fit results for the scaling dimensions and operator product expansion coefficients for the free operators \(\epsilon\) (\(0^+\)) and \(T\)  (\(2^+\)) as a function of the squared lattice spacing. The green results are taken at fixed finite temperature, \(L_t/L=16\), while the grey points in (c) and (d) are extrapolations to zero temperature \(L_t/L\rightarrow \infty\). Extrapolations to the continuum (orange fit functions, blue squares at \(a/R=0\)) are shown to compare with the exact continuum results at zero temperature (black dashed line).}
    \label{fig:free}
\end{figure*}

In particular, we calculated the finite temperature antipodal four-point function for the free theory for the lattice refinements $ L\in \{4,6,8,12,16,24,32\} $ and different cylinder lengths $ L_t $ such that $ L_t / L $ was between 4 and 16. Projecting this data on the Legendre polynomials, we obtained the expansion coefficients \(c^{(free)}_j(t)\) for different values of \(L\) and \(L_t\). The data for the different \(c^{(free)}_j(t)\) is also available in the Zenodo repository associated with this project \cite{Zenodo}. For the two coefficients with \(j=0\) and \(j=2\) also considered in the interacting theory, we then performed multi-exponential fits of the form 
\begin{equation}
    c^{fit}_j(t) = \sum_{k} a_j^{k} \left(e^{-E_j^{k} t a_t/R} + e^{-E_j^{k} (L_t-t) a_t/R} \right)
\end{equation}
in order to extract the leading coefficients and scaling exponents. In these fits, the finite temperature is only taken into account via the addition of the exponentials \(e^{-E_k (L_t-t) a_t/R}\), just like it was done in the interacting theory. We then used a model averaging procedure similar to what was described for the interacting theory in section \ref{sec:interacting} to average over fits to different \(t\)-ranges and, in this free case, different numbers of included exponentials. 

Comparing Eq.~\eqref{eq:cfree} to \eqref{eq:expansion_coeff}, we could calculate the OPE coefficients \(f^{(free)}_{\sigma \sigma \epsilon}\) and \(f^{(free)}_{\sigma \sigma T}\) for the operators \(\epsilon\) and \(T\) of the free CFT from the leading fit parameters \(a_j^0\) and \(E_j^0\) via 
\begin{align}
    f^{(free)2}_{\sigma \sigma \epsilon} &= \frac{a_0^{0}}{4^{E_0^0}} \nonumber \\
    f^{(free)2}_{\sigma \sigma T} &= \frac{3}{8 \cdot 4^{E_2^0}} \left( a_2^{0} - \frac{4 E_0^0}{3 \left( E_0^0 + 1 \right) } a_0^{0} \right)  \label{eqn: f_phiphi}
\end{align}
Our lattice results for these OPE coefficients as well as the scaling dimensions \(\Delta_{\epsilon}^{(free)}\) and \(\Delta_{T}^{(free)}\) corresponding to the leading exponents in \(c_0\) and \(c_2\), \(E_0^0\) and \(E_2^0\), respectively, can be seen in Fig.~\ref{fig:free}. We plot them as a function of \(\left(\frac{a}{R}\right)^2\), which we know to be the leading scaling behavior of lattice artifacts in the free case. The error bars stem from the fitting errors. Note that we do not show the results for \(L=4\) due to the fact that for such small lattice refinements, the fitting errors are several orders of magnitudes higher than for the other refinements.

\begin{table*}[]
    \centering
    \begin{tabular}{cccc}
    \hline \hline
        Quantity & Exact value & \(L_t / L = 16\) &  \(L_t / L \rightarrow \infty \) \\
        \hline
        \(\Delta_{\epsilon}^{(free)}\) & 1 & 1.0000073(94) & - \\
        \(\Delta_{T}^{(free)}\) & 3 & 3.000030(35) & - \\
        \(f^{(free)2}_{\sigma \sigma \epsilon}\) & 2 & 1.998527(93) & 2.00014(15) \\
        \(f^{(free)2}_{\sigma \sigma T}\) & 0.09375 & 0.093681(16) & 0.093769(24) \\
    \hline \hline
    \end{tabular}
    \caption{Extrapolation results for the free theory compared to the zero-temperature exact values (compare Eq.~\eqref{eq:cfree}). For the OPE coefficients \(f^2_{\sigma \sigma \epsilon}\) and \(f^2_{\sigma \sigma T}\) we list the continuum extrapolations obtained from the data with \(L_t/L=16\) fixed, as well as for data which had been extrapolated to zero temperature, \(L_t/L\rightarrow \infty\), prior to the continuum extrapolation \(a/R \rightarrow 0\).}
    \label{tab:free}
\end{table*}

The results in green are taken at finite temperature with fixed \(L_t / L = 16\) just as in the interacting theory. Extrapolating these to the continuum by fitting $ f = c_0 + c_1 \left(\frac{a}{R}\right)^2 $ leads to good agreement with the analytic values in the case of the scaling dimensions, which we expect to be \(\Delta_{\epsilon}^{(free)}=1\) and \(\Delta_{T}^{(free)}=3\) as \(a/R \rightarrow 0\). 

The OPE coefficients, however, are affected by finite temperature effects. Thus, for the fixed finite ratio \(L_t / L = 16\) corresponding to a finite temperature, the extrapolations to \(a/R \rightarrow 0 \) with \(f^{(free)2}_{\sigma \sigma \epsilon} = 1.998527(93)\) and \(f^{(free)2}_{\sigma \sigma T} = 0.093681(16)\) slightly miss the zero temperature continuum expectations from Eq.~\eqref{eq:cfree} corresponding to \(f^{(free)2}_{\sigma \sigma \epsilon} = 2\) and \(f^{(free)2}_{\sigma \sigma T} = 0.09375\). However, these deviations due to wraparound effects only amount to 0.07\%. 

We can eliminate this systematic error by carrying out a double limiting procedure, first extrapolating the lattice results to zero temperature $ L_t/L \rightarrow \infty $ via fit function $ f(L_t) = f + p_1 \ e^{- p_2 L_t } $ for each \(L\), yielding the grey data in Fig.~\ref{fig:free}, and then taking the limit \(a\rightarrow 0\). Fig.~\ref{fig:free} shows that if we perform this procedure, our results perfectly agree with the continuum zero temperature expectations. 

From these results, we can draw two main conclusions. Firstly, as we know the scaling with the lattice spacing exactly, we can perfectly extrapolate the free theory data calculated on our simplicial lattices data to remove the IR and UV cutoff, yielding the free continuum values. This might be an incentive to study the cutoff dependence of the interacting theory to higher than just leading order in order to improve extrapolations.

Secondly, there are wraparound effects influencing our data if we only carry out calculations at finite T (finite \(L_t/L\)) and compare to \(T\rightarrow 0\) ($ L_t/L \rightarrow \infty $) continuum expectations. However, the errors associated with these effects are around two orders of magnitude smaller than the deviations of our OPE coefficient results from the bootstrap values that we see in the interacting theory. Thus, we conclude that the improvement from carrying out a limit $ L_t/L \rightarrow \infty $ in the interacting theory would be negligible compared to the effect of the systematic errors in the interacting theory that we could not study with the free case. 

\section{Conclusion and Outlook}
The Quantum Finite Element program seeks to extend
Euclidean Monte Carlo lattice methods to non-perturbative quantum field theory on curved manifolds. In particular, the present goal is to demonstrate its advantage when applied to radial quantization for $d \ge 3 $ conformal field theory.
For the prototype lattice $\phi^4$ example, we have demonstrated that we can extract scaling dimensions and OPE coefficients for leading as well as subleading operators of the 3d Ising CFT from the four-point function on a \({\mathbb R}\times {\mathbb S}^d\) manifold. In particular, this is the first time, to our knowledge, that the OPE coefficient \(f_{\sigma \sigma T}\) has been determined from lattice Monte Carlo calculations.

Continuum extrapolations of our lattice simulations for the scaling dimensions of operators \(\epsilon\), \(T\) as well as the subleading operator \(\epsilon'\) agree with the values obtained with the conformal bootstrap (and the exact value in the case of the energy-momentum tensor) within the statistical and fitting errors. Extrapolations for the OPE coefficients \(f_{\sigma \sigma \epsilon}\) and \(f_{\sigma \sigma T}\) deviate more from the bootstrap values than is covered by these errors so that also our value for the central charge is significantly too low in violation of the bootstrap lower bound~\cite{ElShowk2014}. However, ratios of these quantities extrapolate to values that are consistent with values obtained by the other two methods that were able to extract both \(f_{\sigma \sigma \epsilon}\) and \(f_{\sigma \sigma T}\), namely the conformal bootstrap and the fuzzy sphere, as is shown in the comparison Tab.~\ref{tab:comparison}.

\begin{table}
    \centering
    \begin{tabular}{cccc}
    \hline \hline
         Quantity & This work & Bootstrap & Fuzzy sphere \\
         \hline
         \(f_{\sigma \sigma \epsilon}^2/f_{\sigma \sigma T}^2\) & 10.30(16) & 10.402(24) & 10.53(23) \\
         \(C_T/C_T^{(free)} \cdot f_{\sigma\sigma \epsilon}^2/\Delta_{\epsilon}^2\)  & 0.528(16) & 0.52478(13) & 0.5310(86)\\
        \hline \hline
    \end{tabular}
    \caption{Comparison of the ratios obtained in this work with the corresponding values obtained with the conformal bootstrap \cite{SimmonsDuffin2016, Reehorst2022} and the Fuzzy sphere \cite{Hu:2023xak}.}
    \label{tab:comparison}
\end{table}

The method employed here follows a sequence of Quantum Finite Element developments. First, there is the 
requirement to introduce UV counterterms in 2d on ${\mathbb S}^2$ and in 3d on ${\mathbb R}\times {\mathbb S}^{d-1}$ to modify the simplicial finite element method (FEM) to reach the Ising critical surface. Next, the importance of the Ricci term in 3d was seen as it dramatically improved the approach to continuum for the scalar two-point function.  While this simplicial QFE action has enabled us to study the OPE expansion, there are clearly additional improvements to pursue, as is common practice in lattice field theory.

Comparison of the free scalar theory on the simplicial lattice \textit{vs.}\ the continuum free CFT reveals two issues. First, the finite temperature (aka IR finite length of the cylinder) is probably not a significant source of systematic error. Second, the extrapolation to zero lattice spacing (\textit{i.e.}\ removing the UV cut-off) for the free CFT with $O(a^2)$ scaling is under good control. For the strongly-interacting CFT, however, the analytic dependence on the cutoff is a more serious problem to address and finite-size scaling analysis offers some guidance.

Within the current QFE framework presented here, there are still some potential sources of systematic error that could be explored further. A slight mistuning of the bare mass
parameter $\mu_0^2$ could potentially lead to a systematic shift in estimated OPE coefficients.
This can only be checked by new calculations at even larger values of $L$.  Similarly, all calculations
were performed at only a single bare $\lambda_0$ with no attempt to extrapolate to the bare lattice $\lambda_0 \to 0$. Also, our determinations of scaling dimensions and OPE coefficients were actually variational estimates limited by our inability to model the contributions of higher primaries: ($\epsilon^{\prime\prime}$,
$T^{\prime\prime}$, $\ell \ge 4$, \textit{etc}). Any of these studies would require a significant investment
in software development for increased parallelization of the code as well as at least an order of magnitude
increase in computational resources.  Before making such an investment, we would like to consider whether
this framework is the best option available.

Beyond cut-off effects, we continue to explore the validity of using perturbative counterterms in super-renormalizable theories.
Recall that our counterterms only correct the mass parameter and do not modify the classical FEM Laplace-Beltrami operator~\cite{StrangFix200805,2005math8341D} in Eq.~(\ref{eq:action_discrete}). However, in a recent study of the 2d Ising model on an affine lattice~\cite{Brower_2023},
equivalent to lattice $\phi^4$ at bare lattice $\lambda_0 \rightarrow \infty$, we demonstrated that the critical theory required a modification of the kinetic term away from the classical FEM operator. This provides yet further evidence that the way to continue using our current simplicial action in Eq.~\ref{eq:QFEaction} with perturbative counterterms
is to fix the dimensional renormalized coupling ($\lambda_R = \mathcal{O}(\lambda_0/a)$) instead of the bare \(\lambda_0\) as we extrapolate to zero lattice spacing. This should
reduce the quantum correction of FEM Laplace-Beltrami operator to $O(a)$.

\section{Acknowledgements}
 The authors are pleased to acknowledge that the computational work reported on in this paper was performed on the Shared Computing Cluster, which is administered by Boston University’s Research Computing Services (\url{www.bu.edu/tech/support/research/}). We thank Casey Berger and Andrew D. Gasbarro, whose calculations this work builds upon. This work is supported by the U.S. Department of Energy (DOE) under Award No. DE-SC0019061 for GTF (Yale), Award No. DE-SC0015845 for RCB, and Award No. DE-SC0019139 for EKO. AMEG acknowledges support from Cusanuswerk Bischöfliche Studienförderung and the Baden-Württemberg Stiftung. This document was prepared by the Quantum Finite Elements collaboration using the resources of the Fermi National Accelerator Laboratory (Fermilab), a U.S. Department of Energy, Office of Science, Office of High Energy Physics HEP User Facility. Fermilab is managed by Fermi Research Alliance, LLC (FRA), acting under Contract No. DE-AC02-07CH11359.
 
\appendix

\section{The free case}
\label{app:free}
In the special case of the free theory, the propagator $G^{(2)}_{\sigma}(t, \cos \theta) \equiv \langle \sigma(t_1, \vec n_1) \sigma(t_2, \vec n_2) \rangle$ for a free scalar field $\sigma(t, \vec n)$ on $\mathbb{R} \times S^2$ is
\begin{equation}
    G^{(2)}_{\sigma}(t, \cos \theta) = \dfrac{2}{\left( 2 \cosh t - 2 \cos \theta \right)^{\Delta_{\sigma}^{(free)}}}    
\end{equation}
where $t=|t_2-t_1|$ and $\cos \theta = \vec n_1 \cdot \vec n_2$. Expanding as a series in Legendre polynomials this is
\begin{equation}
\label{eq:free_2pt}
    G^{(2)}_{\sigma}(t, \cos \theta) = \sum_{j = 0}^{\infty} 2 e^{-(j + 1/2)t} P_{j}(\cos \theta) \;.
\end{equation}
where we have substituted the scaling exponent $\Delta_{\sigma}^{(free)}=1/2$.

The free antipodal four-point function is simply a sum of $s$, $t$, and $u$-channels. Following the same notation used above this gives
\begin{multline}
    G^{(4)}_{\sigma}(t,\cos \theta) = G^{(2)}_{\sigma}(0,-1)^2 + G^{(2)}_{\sigma}(t,\cos \theta)^2 \\+ G^{(2)}_{\sigma}(t,-\cos \theta)^2 \;.
\end{multline}
Note that we have conveniently normalized the 2-point function (\ref{eq:free_2pt}) so that the $s$-channel diagram is 1. Thus, the antipodal 4-point function in this normalization already corresponds to the conformally invariant amplitude \(g\). Inserting Eq. (\ref{eq:free_2pt}) we obtain the somewhat complicated expression
\begin{multline}
    G^{(4)}_{\sigma}(t,\cos \theta) = 1 + 4 \left( \sum_{j = 0}^{\infty} e^{-(j + 1/2)t} P_{\ell}(\cos \theta) \right)^2 \\+ 4 \left( \sum_{j = 0}^{\infty} e^{-(j + 1/2)t} P_{j}(-\cos \theta) \right)^2
\end{multline}
which can again be expressed as a series in Legendre polynomials
\begin{equation}
\label{eq:free_4pt}
    G^{(4)}_{\sigma}(t,\cos \theta) = 1 + \sum_{j=0}^{\infty} c^{(free)}_{j}(t) P_{j}(\cos \theta)
\end{equation}
where all of the odd $j$ coefficients are zero, and the even $j$ coefficients are
\begin{equation}
    c^{(free)}_{j}(t) = 8 (2 j + 1) \sum_{n=0}^{\infty} e^{-(j+2n+1)t} \dfrac{2^{j}(j+n)!(2 n-1)!!}{n!(2 j + 2 n+1)!!}.
    \label{eq:Cl_free}
\end{equation}
As required, we recover the partial wave expansion for a free scalar four-point function.

Finally, we consider corrections to (\ref{eq:free_4pt}) due to wraparound effects on a periodic cylinder with length $L_t$ (i.e. finite temperature effects). In the free case, the thermal 2-point function $\tilde G^{(2)}_{\sigma}$ in the interval $0\leq t \leq L_t$ can be expressed as a sum over an infinite set of image charges.
\begin{multline}
    \label{eq:FiniteT}
    \tilde G^{(2)}_{\sigma}(t,\cos \theta) = \sum_{n=-\infty}^{\infty} G^{(2)}_{\phi}(t + n L_t, \cos \theta) \\
    \\= 4 \sum_{j=0}^{\infty} \dfrac{\cosh\left[(\Delta_{\sigma}^{(free)} + j)(t - L_t / 2)\right]}{\sinh\left[(\Delta_{\sigma}^{(free)} + j) L_t / 2 \right]} P_{j}(\cos \theta)
\end{multline}

The four-point function can be evaluated as before, and to lowest order, the finite temperature Legendre coefficients are related to the zero-temperature coefficients by $\tilde c^{(free)}_{j}(t) \simeq c^{(free)}_{j}(t) + c^{(free)}_{j}(L_t-t)$. The leading corrections occur near $t=L_t/2$, giving
\begin{equation}
    \delta \tilde c^{(free)}_{0}(t \sim L_t/2) \simeq \dfrac{8}{\sinh(\Delta_{\epsilon}^{(free)}L_t / 2)}
\end{equation}
for $j = 0$ and
\begin{equation}
    \delta \tilde c^{(free)}_{j}(t \sim L_t / 2) \simeq 16 \dfrac{\cosh\left[(\Delta_{\epsilon}^{(free)}+ j - 1)(t - L_t / 2)\right]}{\sinh\left[(\Delta_{\epsilon}^{(free)} + j)L_t / 2\right]}
\end{equation}
for all even $j \ge 2$.

\section{Finite-size scaling analysis for the interacting theory}
\label{sec:fss}
To determine which fit function to use for extrapolating the lattice results to the continuum \(a/R\rightarrow 0\), we carried out a finite-size scaling analysis based on Ref.~\cite{Blote:1995zik}. 

\subsection{Scaling of the free energy}

The free energy is defined as 
\begin{equation}
    \mathcal{F}(g_{\sigma}, g_{\epsilon}, \{g_{\omega}, \cdots\}, a/L)  = \log Z
\end{equation}
where
\begin{equation}
    Z = \int D\phi e^{-S+h\cdot\phi}.
\end{equation}
Here, \(g_{\sigma}, g_{\epsilon}\) are relevant and \(g_{\omega}, \cdots\) irrelevant couplings of the action near the Wilson-Fisher fixed point, not the bare couplings \(h, \mu_0, \lambda_0\) in the action. \(a\) is the lattice spacing, \(L\) is the characteristic  ``finite size'' of our system, and \(a/L\) a dimensionless ratio. \\

If we change the scale by a factor \(\lambda\), the free energy renormalizes as follows:
\begin{align}
    \mathcal{F}(g_{\sigma}, &g_{\epsilon}, \{g_{\omega}, \cdots\}, a/L) \\
    = &\lambda^{-d} F(\lambda^{y_{\sigma}} g_{\sigma}, \lambda^{y_{\epsilon}} g_{\epsilon}, \{\lambda^{y_{\omega}}g_{\omega}, \cdots\}, \lambda a/L) + G(g_{\sigma}, g_{\epsilon}) \nonumber
\end{align}
Here, \(F\) is the scaling part of \(\mathcal{F}\), whereas \(G\) is regular. The exponents \(y_{\mathcal{O}}\) are related to the scaling dimensions of the respective operators via \(y_{\mathcal{O}} = d-\Delta_{O}\)

\subsection{\texorpdfstring{$n$}{n}-point functions}
We calculate connected $n$-point functions of the scalar field $\phi$ from \(\mathcal{F}\) by taking derivatives with respect to the (bare) external field \(h\) and subsequently setting \(h=0\):
\begin{align}
    \left. \langle \phi(x_1) \cdots \phi(x_n) \rangle \right\vert_{conn.} = \left. \frac{\delta}{\delta h(x_1)} \cdots \frac{\delta}{\delta h(x_n)} \mathcal{F} \right\vert_{h=0}
    \label{eq:npt}
\end{align}

Rewriting the derivatives with respect to \(g_{\sigma}\), we get for the two-point function
\begin{align}
\begin{split}
     \langle \phi(x_1) &\phi(x_2) \rangle \vert_{conn.} \\
     &\left. = \int_{x, y} \frac{\delta^2 \mathcal{F}}{\delta g_{\sigma}(x) \delta g_{\sigma}(y)} \frac{\delta g_{\sigma}(x)}{\delta h(x_1)} \frac{\delta g_{\sigma}(y)}{\delta h(x_2)} \right\vert_{h=0}
    \label{eq:2pt}
\end{split}
\end{align}
where we have used that \(g_{\sigma}\) is an odd function of \(h\) 
\begin{equation}
 g_{\sigma} = h\alpha_1 + h^3\frac{\alpha_3}{3!} + h^5\frac{\alpha_5}{5!} + \cdots   
\end{equation}
such that terms like \(\frac{\delta^2 g_{\sigma}(x)}{\delta h(x_1)\delta h(x_2)}\) vanish.
For the four-point function, we get 
\begin{align*}
     \langle \phi(&x_1) \phi(x_2) \phi(x_3) \phi(x_4) \rangle \vert_{conn.} \\ 
    =& \int\limits_{x, y, u, w} \frac{\delta^4 \mathcal{F}}{\delta g_{\sigma}(x) \delta g_{\sigma}(y) g_{\sigma}(u) \delta g_{\sigma}(w)}   \\
    & \qquad \qquad \times \frac{\delta g_{\sigma}(x)}{\delta h(x_1)} \frac{\delta g_{\sigma}(y)}{\delta h(x_2)} \frac{\delta g_{\sigma}(u)}{\delta h(x_3)} \frac{\delta g_{\sigma}(w)}{\delta h(x_4)} \\
    +& \int\limits_{x, y} \frac{\delta^2 \mathcal{F}}{\delta g_{\sigma}(x) \delta g_{\sigma}(y)} \biggl( \frac{\delta g_{\sigma}(x)}{\delta h(x_1)} \frac{\delta^3 g_{\sigma}(y)}{\delta h(x_2) \delta h(x_3) \delta h(x_4)} \\
    & \qquad \qquad + \frac{\delta g_{\sigma}(x)}{\delta h(x_2)} \frac{\delta^3 g_{\sigma}(y)}{\delta h(x_1) \delta h(x_3) \delta h(x_4)} \\ 
    & \qquad \qquad +\frac{\delta g_{\sigma}(x)}{\delta h(x_3)} \frac{\delta^3 g_{\sigma}(y)}{\delta h(x_1) \delta h(x_2) \delta h(x_4)} \\
    & \qquad \qquad + \frac{\delta g_{\sigma}(x)}{\delta h(x_4)} \frac{\delta^3 g_{\sigma}(y)}{\delta h(x_1) \delta h(x_2) \delta h(x_3)} \biggl)
\end{align*}

Derivatives of the free energy with respect to \(g_{\sigma}\) in the limit of \(h \rightarrow 0\) and thus \(g_{\sigma}\rightarrow 0\) scale as 

\begin{align}
    &\frac{\partial^k \mathcal{F}}{\partial g_{\sigma}^k} = \mathcal{F}^{(k)}(g_{\epsilon},  \{g_{\omega}, \cdots\}, a/L) \\
    & = \left(\frac{L}{a}\right)^{ky_{\sigma}-d} F^{(k)}\left(\left(\frac{L}{a}\right)^{y_{\epsilon}} g_{\epsilon}, \left\{\left(\frac{L}{a}\right)^{y_{\omega}}g_{\omega}, \cdots\right\}, 1\right) \nonumber \\
    &\quad + G^{(k)}(g_{\epsilon}) \nonumber
\end{align}

Near the critical point, we can expand these derivatives in Taylor series as follows:
\begin{align}
    &F^{(k)} \\
    & = a_{k0} + a_{k1}(g_{\epsilon}-g_{\epsilon}^*)\left(\frac{L}{a}\right)^{y_{\epsilon}} + a_{k2}(g_{\epsilon}-g_{\epsilon}^*)^2\left(\frac{L}{a}\right)^{2y_{\epsilon}} + \cdots \nonumber \\
    &\quad + b_{k1} (g_{\omega}-g_{\omega}^*)\left(\frac{L}{a}\right)^{y_{\omega}} + b_{k2} (g_{\omega}-g_{\omega}^*)^2\left(\frac{L}{a}\right)^{2y_{\omega}} + \cdots \nonumber
\end{align}
and 
\begin{align}
    G^{(k)} = c_{k0} + c_{k1}(g_{\epsilon}-g_{\epsilon}^*) + c_{k2}(g_{\epsilon}-g_{\epsilon}^*)^2 + \cdots
\end{align}
where \(g^*\) denotes the critical values of the couplings at the Wilson-Fisher fixed point. 

Putting everything together, we see that the two-point function scales as 
\begin{align}
    \left . \langle \phi\phi \rangle \right \vert_{conn.} \sim \left(\frac{L}{a}\right)^{2y_{\sigma}-2d} \alpha_1^2 \left[ F^{(2)} +  \left(\frac{L}{a}\right)^{d-2y_{\sigma}} G^{(2)}\right] 
\end{align}
near the critical point, whereas the connected four-point function scales as 
\begin{align}
    & \left . \langle \phi\phi\phi\phi \rangle \right \vert_{conn.}  \sim  \left(\frac{L}{a}\right)^{4y_{\sigma}-4d}\alpha_1^4 \left[ F^{(4)} +  \left(\frac{L}{a}\right)^{d-4y_{\sigma}} G^{(4)} \right] \nonumber \\
    & + 4 \left(\frac{L}{a}\right)^{2y_{\sigma}-4d}\alpha_1\alpha_3  \left[ F^{(2)} +  \left(\frac{L}{a}\right)^{d-2y_{\sigma}} G^{(2)} \right] 
\end{align}
By tuning \(\mu_0^2\) in our action to the critical surface, we eliminate all terms \(\propto (g_{\epsilon}-g_{\epsilon}^*=0)\), which diverge as \(L\rightarrow\infty\). The remaining terms then yield 
\begin{align}
\label{eq:2pt_scale}
   & \left .  \langle \phi\phi \rangle \right \vert_{conn.} \\
   & \sim \left(\frac{L}{a}\right)^{-2\Delta_{\sigma}} \alpha_1^2 \left[ a_{20} + b_{21} (g_{\omega}-g_{\omega}^*)\left(\frac{L}{a}\right)^{d-\Delta_{\omega'}} + \cdots \right.\nonumber \\
   & \left. \quad + \left(\frac{L}{a}\right)^{2\Delta_{\sigma}-d} c_{20} \right] \nonumber
\end{align}
and
\begin{align}
    \label{eq:4pt_scale}
    & \left. \langle \phi\phi\phi\phi \rangle \right \vert_{conn.}  \\ 
    & \sim \left(\frac{L}{a}\right)^{-4\Delta_{\sigma}}\alpha_1^4 \left[ a_{40} + b_{41} (g_{\omega}-g_{\omega}^*)\left(\frac{L}{a}\right)^{d-\Delta_{\omega'}} +  \cdots  \right . \nonumber \\
    &+ \left . \left(\frac{L}{a}\right)^{-3d+4\Delta_{\sigma}} c_{40} \right] + 4 \left(\frac{L}{a}\right)^{-2\Delta_{\sigma}-2d} \alpha_1\alpha_3 \nonumber \\
    &\,\, \times \left[ a_{20} + b_{21} (g_{\omega}-g_{\omega}^*)\left(\frac{L}{a}\right)^{d-\Delta_{\omega'}} + \cdots + \left(\frac{L}{a}\right)^{2\Delta_{\sigma}-d} c_{20} \right] \nonumber 
\end{align}
We now have to take the ratio \(\frac{\left.\langle \phi\phi\phi\phi \rangle\right\vert_{conn.}}{\langle \phi\phi \rangle \langle \phi\phi \rangle} \) of Eq.~(\ref{eq:4pt_scale}) and Eq.~(\ref{eq:2pt_scale}) and expand in \(a/L\). Using that for the \(d=3\) Ising model, the leading irrelevant operator \(\omega\) corresponds to the conformal primary \(\epsilon'\) and comparing with the values for \(\Delta_{\sigma}\) and \(\Delta_{\epsilon'}\) in \cite{SimmonsDuffin2016}, we find that to leading order
\begin{align}
\label{eq:4pt_FSS}
    \frac{\left.\langle \phi\phi\phi\phi \rangle\right\vert_{conn.}}{\langle \phi\phi \rangle \langle \phi\phi \rangle} &= g(u, v)-1 \nonumber \\
    & \sim C_0 + C_1 \left(\frac{a}{L}\right)^{\Delta_{\epsilon'}-3} + \cdots
\end{align}
such that the leading finite-size scaling exponent is \(\Delta_{\epsilon'}-3\approx0.83\). In our case, the characteristic "finite size" of our system is the radius of the sphere \(R\). Thus, we expect the OPE coefficients from our fits at finite lattice spacing to receive corrections 
\begin{equation}
    f_{\sigma\sigma\mathcal{O}}^2(\frac{a}{R}) =  f_{\sigma\sigma\mathcal{O}}^2 + c_{\mathcal{O}}' \left(\frac{a}{R}\right)^{\Delta_{\epsilon'}-3} + \cdots
\end{equation}
to leading order. Taking the logarithm, we find that also 
\begin{equation}
    \Delta_{\mathcal{O}}(\frac{a}{R}) =  \Delta_{\mathcal{O}} + c_{\mathcal{O}} \left(\frac{a}{R}\right)^{\Delta_{\epsilon'}-3} + \cdots 
\end{equation}

Note, $C_1$ in Eq.~(\ref{eq:4pt_FSS}) as the leading correction to scaling is a function of the irrelevant couplings
and its value can be set to zero by tuning the bare couplings \cite{Hasenbusch:1998gh}, thereby increasing the rate of convergence of the discrete theory to the continuum limit.
This is often referred to as non-perturbative improvement of the discrete action.

\bibliography{proposal}{}

\begin{thebibliography}{48}%
\makeatletter
\providecommand \@ifxundefined [1]{%
 \@ifx{#1\undefined}
}%
\providecommand \@ifnum [1]{%
 \ifnum #1\expandafter \@firstoftwo
 \else \expandafter \@secondoftwo
 \fi
}%
\providecommand \@ifx [1]{%
 \ifx #1\expandafter \@firstoftwo
 \else \expandafter \@secondoftwo
 \fi
}%
\providecommand \natexlab [1]{#1}%
\providecommand \enquote  [1]{``#1''}%
\providecommand \bibnamefont  [1]{#1}%
\providecommand \bibfnamefont [1]{#1}%
\providecommand \citenamefont [1]{#1}%
\providecommand \href@noop [0]{\@secondoftwo}%
\providecommand \href [0]{\begingroup \@sanitize@url \@href}%
\providecommand \@href[1]{\@@startlink{#1}\@@href}%
\providecommand \@@href[1]{\endgroup#1\@@endlink}%
\providecommand \@sanitize@url [0]{\catcode `\\12\catcode `\$12\catcode
  `\&12\catcode `\#12\catcode `\^12\catcode `\_12\catcode `\%12\relax}%
\providecommand \@@startlink[1]{}%
\providecommand \@@endlink[0]{}%
\providecommand \url  [0]{\begingroup\@sanitize@url \@url }%
\providecommand \@url [1]{\endgroup\@href {#1}{\urlprefix }}%
\providecommand \urlprefix  [0]{URL }%
\providecommand \Eprint [0]{\href }%
\providecommand \doibase [0]{https://doi.org/}%
\providecommand \selectlanguage [0]{\@gobble}%
\providecommand \bibinfo  [0]{\@secondoftwo}%
\providecommand \bibfield  [0]{\@secondoftwo}%
\providecommand \translation [1]{[#1]}%
\providecommand \BibitemOpen [0]{}%
\providecommand \bibitemStop [0]{}%
\providecommand \bibitemNoStop [0]{.\EOS\space}%
\providecommand \EOS [0]{\spacefactor3000\relax}%
\providecommand \BibitemShut  [1]{\csname bibitem#1\endcsname}%
\let\auto@bib@innerbib\@empty
\bibitem [{\citenamefont {Andrews}(1869)}]{Andrews1869}%
  \BibitemOpen
  \bibfield  {author} {\bibinfo {author} {\bibfnamefont {T.}~\bibnamefont
  {Andrews}},\ }\bibfield  {title} {\bibinfo {title} {The bakerian lecture: On
  the continuity of the gaseous and liquid states of matter},\ }\href
  {http://www.jstor.org/stable/109009} {\bibfield  {journal} {\bibinfo
  {journal} {Philosophical Transactions of the Royal Society of London}\
  }\textbf {\bibinfo {volume} {159}},\ \bibinfo {pages} {575} (\bibinfo {year}
  {1869})}\BibitemShut {NoStop}%
\bibitem [{\citenamefont {Wilson}(1971)}]{Wilson1971}%
  \BibitemOpen
  \bibfield  {author} {\bibinfo {author} {\bibfnamefont {K.~G.}\ \bibnamefont
  {Wilson}},\ }\bibfield  {title} {\bibinfo {title} {Renormalization group and
  critical phenomena. i. renormalization group and the kadanoff scaling
  picture},\ }\href {https://doi.org/10.1103/PhysRevB.4.3174} {\bibfield
  {journal} {\bibinfo  {journal} {Phys. Rev. B}\ }\textbf {\bibinfo {volume}
  {4}},\ \bibinfo {pages} {3174} (\bibinfo {year} {1971})}\BibitemShut
  {NoStop}%
\bibitem [{\citenamefont {Parola}\ and\ \citenamefont
  {Reatto}(1995)}]{Parola1995}%
  \BibitemOpen
  \bibfield  {author} {\bibinfo {author} {\bibfnamefont {A.}~\bibnamefont
  {Parola}}\ and\ \bibinfo {author} {\bibfnamefont {L.}~\bibnamefont
  {Reatto}},\ }\bibfield  {title} {\bibinfo {title} {Liquid state theories and
  critical phenomena},\ }\href {https://doi.org/10.1080/00018739500101536}
  {\bibfield  {journal} {\bibinfo  {journal} {Advances in Physics}\ }\textbf
  {\bibinfo {volume} {44}},\ \bibinfo {pages} {211} (\bibinfo {year} {1995})},\
  \Eprint {https://arxiv.org/abs/https://doi.org/10.1080/00018739500101536}
  {https://doi.org/10.1080/00018739500101536} \BibitemShut {NoStop}%
\bibitem [{\citenamefont {Anisimov}(1991)}]{Anisimov1991}%
  \BibitemOpen
  \bibfield  {author} {\bibinfo {author} {\bibfnamefont {M.}~\bibnamefont
  {Anisimov}},\ }\href {https://books.google.de/books?id=3QaTMcaWDxQC} {\emph
  {\bibinfo {title} {Critical Phenomena in Liquids and Liquid Crystals}}}\
  (\bibinfo  {publisher} {Gordon and Breach Science Publishers},\ \bibinfo
  {year} {1991})\BibitemShut {NoStop}%
\bibitem [{\citenamefont {Pelissetto}\ and\ \citenamefont
  {Vicari}(2002)}]{Pelissetto2000}%
  \BibitemOpen
  \bibfield  {author} {\bibinfo {author} {\bibfnamefont {A.}~\bibnamefont
  {Pelissetto}}\ and\ \bibinfo {author} {\bibfnamefont {E.}~\bibnamefont
  {Vicari}},\ }\bibfield  {title} {\bibinfo {title} {{Critical phenomena and
  renormalization group theory}},\ }\href
  {https://doi.org/10.1016/S0370-1573(02)00219-3} {\bibfield  {journal}
  {\bibinfo  {journal} {Phys. Rept.}\ }\textbf {\bibinfo {volume} {368}},\
  \bibinfo {pages} {549} (\bibinfo {year} {2002})},\ \Eprint
  {https://arxiv.org/abs/cond-mat/0012164} {arXiv:cond-mat/0012164}
  \BibitemShut {NoStop}%
\bibitem [{\citenamefont {Swendsen}\ and\ \citenamefont
  {Wang}(1987)}]{Swendsen1987}%
  \BibitemOpen
  \bibfield  {author} {\bibinfo {author} {\bibfnamefont {R.~H.}\ \bibnamefont
  {Swendsen}}\ and\ \bibinfo {author} {\bibfnamefont {J.-S.}\ \bibnamefont
  {Wang}},\ }\bibfield  {title} {\bibinfo {title} {{Nonuniversal critical
  dynamics in Monte Carlo simulations}},\ }\href
  {https://doi.org/10.1103/PhysRevLett.58.86} {\bibfield  {journal} {\bibinfo
  {journal} {Phys. Rev. Lett.}\ }\textbf {\bibinfo {volume} {58}},\ \bibinfo
  {pages} {86} (\bibinfo {year} {1987})}\BibitemShut {NoStop}%
\bibitem [{\citenamefont {Wolff}(1989)}]{Wolff1989}%
  \BibitemOpen
  \bibfield  {author} {\bibinfo {author} {\bibfnamefont {U.}~\bibnamefont
  {Wolff}},\ }\bibfield  {title} {\bibinfo {title} {{Collective Monte Carlo
  Updating for Spin Systems}},\ }\href
  {https://doi.org/10.1103/PhysRevLett.62.361} {\bibfield  {journal} {\bibinfo
  {journal} {Phys. Rev. Lett.}\ }\textbf {\bibinfo {volume} {62}},\ \bibinfo
  {pages} {361} (\bibinfo {year} {1989})}\BibitemShut {NoStop}%
\bibitem [{\citenamefont {Hasenbusch}(2010)}]{Hasenbusch2010}%
  \BibitemOpen
  \bibfield  {author} {\bibinfo {author} {\bibfnamefont {M.}~\bibnamefont
  {Hasenbusch}},\ }\bibfield  {title} {\bibinfo {title} {{Finite size scaling
  study of lattice models in the three-dimensional Ising universality class}},\
  }\href {https://doi.org/10.1103/PhysRevB.82.174433} {\bibfield  {journal}
  {\bibinfo  {journal} {Phys. Rev. B}\ }\textbf {\bibinfo {volume} {82}},\
  \bibinfo {pages} {174433} (\bibinfo {year} {2010})},\ \Eprint
  {https://arxiv.org/abs/1004.4486} {arXiv:1004.4486 [cond-mat.stat-mech]}
  \BibitemShut {NoStop}%
\bibitem [{\citenamefont {Polyakov}(1970)}]{Polyakov:1970xd}%
  \BibitemOpen
  \bibfield  {author} {\bibinfo {author} {\bibfnamefont {A.~M.}\ \bibnamefont
  {Polyakov}},\ }\bibfield  {title} {\bibinfo {title} {{Conformal symmetry of
  critical fluctuations}},\ }\href@noop {} {\bibfield  {journal} {\bibinfo
  {journal} {JETP Lett.}\ }\textbf {\bibinfo {volume} {12}},\ \bibinfo {pages}
  {381} (\bibinfo {year} {1970})}\BibitemShut {NoStop}%
\bibitem [{\citenamefont {Rose}\ \emph {et~al.}(2022)\citenamefont {Rose},
  \citenamefont {Pagani},\ and\ \citenamefont {Dupuis}}]{Rose2022}%
  \BibitemOpen
  \bibfield  {author} {\bibinfo {author} {\bibfnamefont {F.}~\bibnamefont
  {Rose}}, \bibinfo {author} {\bibfnamefont {C.}~\bibnamefont {Pagani}},\ and\
  \bibinfo {author} {\bibfnamefont {N.}~\bibnamefont {Dupuis}},\ }\bibfield
  {title} {\bibinfo {title} {Operator product expansion coefficients from the
  nonperturbative functional renormalization group},\ }\href
  {https://doi.org/10.1103/PhysRevD.105.065020} {\bibfield  {journal} {\bibinfo
   {journal} {Phys. Rev. D}\ }\textbf {\bibinfo {volume} {105}},\ \bibinfo
  {pages} {065020} (\bibinfo {year} {2022})}\BibitemShut {NoStop}%
\bibitem [{\citenamefont {Hasenbusch}(2018)}]{Hasenbusch2018}%
  \BibitemOpen
  \bibfield  {author} {\bibinfo {author} {\bibfnamefont {M.}~\bibnamefont
  {Hasenbusch}},\ }\bibfield  {title} {\bibinfo {title} {Two- and three-point
  functions at criticality: Monte carlo simulations of the improved
  three-dimensional blume-capel model},\ }\href
  {https://doi.org/10.1103/PhysRevE.97.012119} {\bibfield  {journal} {\bibinfo
  {journal} {Phys. Rev. E}\ }\textbf {\bibinfo {volume} {97}},\ \bibinfo
  {pages} {012119} (\bibinfo {year} {2018})}\BibitemShut {NoStop}%
\bibitem [{\citenamefont {Herdeiro}(2017)}]{Herdeiro2017}%
  \BibitemOpen
  \bibfield  {author} {\bibinfo {author} {\bibfnamefont {V.}~\bibnamefont
  {Herdeiro}},\ }\bibfield  {title} {\bibinfo {title} {Numerical estimation of
  structure constants in the three-dimensional ising conformal field theory
  through markov chain uv sampler},\ }\href
  {https://doi.org/10.1103/PhysRevE.96.033301} {\bibfield  {journal} {\bibinfo
  {journal} {Phys. Rev. E}\ }\textbf {\bibinfo {volume} {96}},\ \bibinfo
  {pages} {033301} (\bibinfo {year} {2017})}\BibitemShut {NoStop}%
\bibitem [{\citenamefont {Caselle}\ \emph {et~al.}(2015)\citenamefont
  {Caselle}, \citenamefont {Costagliola},\ and\ \citenamefont
  {Magnoli}}]{Caselle2015}%
  \BibitemOpen
  \bibfield  {author} {\bibinfo {author} {\bibfnamefont {M.}~\bibnamefont
  {Caselle}}, \bibinfo {author} {\bibfnamefont {G.}~\bibnamefont
  {Costagliola}},\ and\ \bibinfo {author} {\bibfnamefont {N.}~\bibnamefont
  {Magnoli}},\ }\bibfield  {title} {\bibinfo {title} {Numerical determination
  of the operator-product-expansion coefficients in the 3d ising model from
  off-critical correlators},\ }\href
  {https://doi.org/10.1103/PhysRevD.91.061901} {\bibfield  {journal} {\bibinfo
  {journal} {Phys. Rev. D}\ }\textbf {\bibinfo {volume} {91}},\ \bibinfo
  {pages} {061901} (\bibinfo {year} {2015})}\BibitemShut {NoStop}%
\bibitem [{\citenamefont {Costagliola}(2016)}]{Costagliola2016}%
  \BibitemOpen
  \bibfield  {author} {\bibinfo {author} {\bibfnamefont {G.}~\bibnamefont
  {Costagliola}},\ }\bibfield  {title} {\bibinfo {title} {Operator product
  expansion coefficients of the 3d ising model with a trapping potential},\
  }\href {https://doi.org/10.1103/PhysRevD.93.066008} {\bibfield  {journal}
  {\bibinfo  {journal} {Phys. Rev. D}\ }\textbf {\bibinfo {volume} {93}},\
  \bibinfo {pages} {066008} (\bibinfo {year} {2016})}\BibitemShut {NoStop}%
\bibitem [{\citenamefont {Cardy}(1985)}]{Cardy:1985lth}%
  \BibitemOpen
  \bibfield  {author} {\bibinfo {author} {\bibfnamefont {J.~L.}\ \bibnamefont
  {Cardy}},\ }\bibfield  {title} {\bibinfo {title} {{Universal amplitudes in
  finite-size scaling: generalisation to arbitrary dimensionality}},\ }\href
  {https://doi.org/10.1088/0305-4470/18/13/005} {\bibfield  {journal} {\bibinfo
   {journal} {J. Phys. A}\ }\textbf {\bibinfo {volume} {18}},\ \bibinfo {pages}
  {L757} (\bibinfo {year} {1985})}\BibitemShut {NoStop}%
\bibitem [{\citenamefont {Fubini}\ \emph {et~al.}(1973)\citenamefont {Fubini},
  \citenamefont {Hanson},\ and\ \citenamefont {Jackiw}}]{Fubini:1972mf}%
  \BibitemOpen
  \bibfield  {author} {\bibinfo {author} {\bibfnamefont {S.}~\bibnamefont
  {Fubini}}, \bibinfo {author} {\bibfnamefont {A.~J.}\ \bibnamefont {Hanson}},\
  and\ \bibinfo {author} {\bibfnamefont {R.}~\bibnamefont {Jackiw}},\
  }\bibfield  {title} {\bibinfo {title} {{New approach to field theory}},\
  }\href {https://doi.org/10.1103/PhysRevD.7.1732} {\bibfield  {journal}
  {\bibinfo  {journal} {Phys.Rev.}\ }\textbf {\bibinfo {volume} {D7}},\
  \bibinfo {pages} {1732} (\bibinfo {year} {1973})}\BibitemShut {NoStop}%
\bibitem [{\citenamefont {Brower}\ \emph {et~al.}(2016)\citenamefont {Brower},
  \citenamefont {Fleming}, \citenamefont {Gasbarro}, \citenamefont {Raben},
  \citenamefont {Tan},\ and\ \citenamefont {Weinberg}}]{Brower:2016moq}%
  \BibitemOpen
  \bibfield  {author} {\bibinfo {author} {\bibfnamefont {R.~C.}\ \bibnamefont
  {Brower}}, \bibinfo {author} {\bibfnamefont {G.}~\bibnamefont {Fleming}},
  \bibinfo {author} {\bibfnamefont {A.}~\bibnamefont {Gasbarro}}, \bibinfo
  {author} {\bibfnamefont {T.}~\bibnamefont {Raben}}, \bibinfo {author}
  {\bibfnamefont {C.-I.}\ \bibnamefont {Tan}},\ and\ \bibinfo {author}
  {\bibfnamefont {E.}~\bibnamefont {Weinberg}},\ }\bibfield  {title} {\bibinfo
  {title} {{Quantum Finite Elements for Lattice Field Theory}},\ }\href
  {https://doi.org/10.22323/1.251.0296} {\bibfield  {journal} {\bibinfo
  {journal} {PoS}\ }\textbf {\bibinfo {volume} {LATTICE2015}},\ \bibinfo
  {pages} {296} (\bibinfo {year} {2016})},\ \Eprint
  {https://arxiv.org/abs/1601.01367} {arXiv:1601.01367 [hep-lat]} \BibitemShut
  {NoStop}%
\bibitem [{\citenamefont {Brower}\ \emph {et~al.}(2018)\citenamefont {Brower},
  \citenamefont {Cheng}, \citenamefont {Weinberg}, \citenamefont {Fleming},
  \citenamefont {Gasbarro}, \citenamefont {Raben},\ and\ \citenamefont
  {Tan}}]{Brower2018}%
  \BibitemOpen
  \bibfield  {author} {\bibinfo {author} {\bibfnamefont {R.~C.}\ \bibnamefont
  {Brower}}, \bibinfo {author} {\bibfnamefont {M.}~\bibnamefont {Cheng}},
  \bibinfo {author} {\bibfnamefont {E.~S.}\ \bibnamefont {Weinberg}}, \bibinfo
  {author} {\bibfnamefont {G.~T.}\ \bibnamefont {Fleming}}, \bibinfo {author}
  {\bibfnamefont {A.~D.}\ \bibnamefont {Gasbarro}}, \bibinfo {author}
  {\bibfnamefont {T.~G.}\ \bibnamefont {Raben}},\ and\ \bibinfo {author}
  {\bibfnamefont {C.-I.}\ \bibnamefont {Tan}},\ }\bibfield  {title} {\bibinfo
  {title} {{Lattice $\phi^4$ field theory on Riemann manifolds: Numerical tests
  for the 2-d Ising CFT on $\mathbb{S}^2$}},\ }\href
  {https://doi.org/10.1103/PhysRevD.98.014502} {\bibfield  {journal} {\bibinfo
  {journal} {Phys. Rev. D}\ }\textbf {\bibinfo {volume} {98}},\ \bibinfo
  {pages} {014502} (\bibinfo {year} {2018})},\ \Eprint
  {https://arxiv.org/abs/1803.08512} {arXiv:1803.08512 [hep-lat]} \BibitemShut
  {NoStop}%
\bibitem [{\citenamefont {Brower}\ \emph {et~al.}(2017)\citenamefont {Brower},
  \citenamefont {Weinberg}, \citenamefont {Fleming}, \citenamefont {Gasbarro},
  \citenamefont {Raben},\ and\ \citenamefont {Tan}}]{PhysRevD.95.114510}%
  \BibitemOpen
  \bibfield  {author} {\bibinfo {author} {\bibfnamefont {R.~C.}\ \bibnamefont
  {Brower}}, \bibinfo {author} {\bibfnamefont {E.~S.}\ \bibnamefont
  {Weinberg}}, \bibinfo {author} {\bibfnamefont {G.~T.}\ \bibnamefont
  {Fleming}}, \bibinfo {author} {\bibfnamefont {A.~D.}\ \bibnamefont
  {Gasbarro}}, \bibinfo {author} {\bibfnamefont {T.~G.}\ \bibnamefont
  {Raben}},\ and\ \bibinfo {author} {\bibfnamefont {C.-I.}\ \bibnamefont
  {Tan}},\ }\bibfield  {title} {\bibinfo {title} {Lattice dirac fermions on a
  simplicial riemannian manifold},\ }\href
  {https://doi.org/10.1103/PhysRevD.95.114510} {\bibfield  {journal} {\bibinfo
  {journal} {Phys. Rev. D}\ }\textbf {\bibinfo {volume} {95}},\ \bibinfo
  {pages} {114510} (\bibinfo {year} {2017})}\BibitemShut {NoStop}%
\bibitem [{\citenamefont {Brower}\ \emph {et~al.}(2021)\citenamefont {Brower},
  \citenamefont {Fleming}, \citenamefont {Gasbarro}, \citenamefont {Howarth},
  \citenamefont {Raben}, \citenamefont {Tan},\ and\ \citenamefont
  {Weinberg}}]{Brower:2020jqj}%
  \BibitemOpen
  \bibfield  {author} {\bibinfo {author} {\bibfnamefont {R.~C.}\ \bibnamefont
  {Brower}}, \bibinfo {author} {\bibfnamefont {G.~T.}\ \bibnamefont {Fleming}},
  \bibinfo {author} {\bibfnamefont {A.~D.}\ \bibnamefont {Gasbarro}}, \bibinfo
  {author} {\bibfnamefont {D.}~\bibnamefont {Howarth}}, \bibinfo {author}
  {\bibfnamefont {T.~G.}\ \bibnamefont {Raben}}, \bibinfo {author}
  {\bibfnamefont {C.-I.}\ \bibnamefont {Tan}},\ and\ \bibinfo {author}
  {\bibfnamefont {E.~S.}\ \bibnamefont {Weinberg}},\ }\bibfield  {title}
  {\bibinfo {title} {{Radial lattice quantization of 3D \ensuremath{\phi}4
  field theory}},\ }\href {https://doi.org/10.1103/PhysRevD.104.094502}
  {\bibfield  {journal} {\bibinfo  {journal} {Phys. Rev. D}\ }\textbf {\bibinfo
  {volume} {104}},\ \bibinfo {pages} {094502} (\bibinfo {year} {2021})},\
  \Eprint {https://arxiv.org/abs/2006.15636} {arXiv:2006.15636 [hep-lat]}
  \BibitemShut {NoStop}%
\bibitem [{\citenamefont {El-Showk}\ \emph {et~al.}(2012)\citenamefont
  {El-Showk}, \citenamefont {Paulos}, \citenamefont {Poland}, \citenamefont
  {Rychkov}, \citenamefont {Simmons-Duffin},\ and\ \citenamefont
  {Vichi}}]{ElShowk2012}%
  \BibitemOpen
  \bibfield  {author} {\bibinfo {author} {\bibfnamefont {S.}~\bibnamefont
  {El-Showk}}, \bibinfo {author} {\bibfnamefont {M.~F.}\ \bibnamefont
  {Paulos}}, \bibinfo {author} {\bibfnamefont {D.}~\bibnamefont {Poland}},
  \bibinfo {author} {\bibfnamefont {S.}~\bibnamefont {Rychkov}}, \bibinfo
  {author} {\bibfnamefont {D.}~\bibnamefont {Simmons-Duffin}},\ and\ \bibinfo
  {author} {\bibfnamefont {A.}~\bibnamefont {Vichi}},\ }\bibfield  {title}
  {\bibinfo {title} {{Solving the 3D Ising Model with the Conformal
  Bootstrap}},\ }\href {https://doi.org/10.1103/PhysRevD.86.025022} {\bibfield
  {journal} {\bibinfo  {journal} {Phys. Rev. D}\ }\textbf {\bibinfo {volume}
  {86}},\ \bibinfo {pages} {025022} (\bibinfo {year} {2012})},\ \Eprint
  {https://arxiv.org/abs/1203.6064} {arXiv:1203.6064 [hep-th]} \BibitemShut
  {NoStop}%
\bibitem [{\citenamefont {El-Showk}\ \emph {et~al.}(2014)\citenamefont
  {El-Showk}, \citenamefont {Paulos}, \citenamefont {Poland}, \citenamefont
  {Rychkov}, \citenamefont {Simmons-Duffin},\ and\ \citenamefont
  {Vichi}}]{ElShowk2014}%
  \BibitemOpen
  \bibfield  {author} {\bibinfo {author} {\bibfnamefont {S.}~\bibnamefont
  {El-Showk}}, \bibinfo {author} {\bibfnamefont {M.~F.}\ \bibnamefont
  {Paulos}}, \bibinfo {author} {\bibfnamefont {D.}~\bibnamefont {Poland}},
  \bibinfo {author} {\bibfnamefont {S.}~\bibnamefont {Rychkov}}, \bibinfo
  {author} {\bibfnamefont {D.}~\bibnamefont {Simmons-Duffin}},\ and\ \bibinfo
  {author} {\bibfnamefont {A.}~\bibnamefont {Vichi}},\ }\bibfield  {title}
  {\bibinfo {title} {{Solving the 3d Ising Model with the Conformal Bootstrap
  II. c-Minimization and Precise Critical Exponents}},\ }\href
  {https://doi.org/10.1007/s10955-014-1042-7} {\bibfield  {journal} {\bibinfo
  {journal} {J. Stat. Phys.}\ }\textbf {\bibinfo {volume} {157}},\ \bibinfo
  {pages} {869} (\bibinfo {year} {2014})},\ \Eprint
  {https://arxiv.org/abs/1403.4545} {arXiv:1403.4545 [hep-th]} \BibitemShut
  {NoStop}%
\bibitem [{\citenamefont {Simmons-Duffin}(2017)}]{SimmonsDuffin2016}%
  \BibitemOpen
  \bibfield  {author} {\bibinfo {author} {\bibfnamefont {D.}~\bibnamefont
  {Simmons-Duffin}},\ }\bibfield  {title} {\bibinfo {title} {{The Lightcone
  Bootstrap and the Spectrum of the 3d Ising CFT}},\ }\href
  {https://doi.org/10.1007/JHEP03(2017)086} {\bibfield  {journal} {\bibinfo
  {journal} {JHEP}\ }\textbf {\bibinfo {volume} {03}},\ \bibinfo {pages}
  {086}},\ \Eprint {https://arxiv.org/abs/1612.08471} {arXiv:1612.08471
  [hep-th]} \BibitemShut {NoStop}%
\bibitem [{\citenamefont {Reehorst}(2022)}]{Reehorst2022}%
  \BibitemOpen
  \bibfield  {author} {\bibinfo {author} {\bibfnamefont {M.}~\bibnamefont
  {Reehorst}},\ }\bibfield  {title} {\bibinfo {title} {Rigorous bounds on
  irrelevant operators in the 3d ising model {CFT}},\ }\bibfield  {journal}
  {\bibinfo  {journal} {Journal of High Energy Physics}\ }\textbf {\bibinfo
  {volume} {2022}},\ \href {https://doi.org/10.1007/jhep09(2022)177}
  {10.1007/jhep09(2022)177} (\bibinfo {year} {2022})\BibitemShut {NoStop}%
\bibitem [{\citenamefont {Gl\"uck}\ \emph {et~al.}(2023)\citenamefont
  {Gl\"uck}, \citenamefont {Fleming}, \citenamefont {Brower}, \citenamefont
  {Ayyar}, \citenamefont {Owen}, \citenamefont {Raben},\ and\ \citenamefont
  {Tan}}]{Gluck:2023zji}%
  \BibitemOpen
  \bibfield  {author} {\bibinfo {author} {\bibfnamefont {A.-M.~E.}\
  \bibnamefont {Gl\"uck}}, \bibinfo {author} {\bibfnamefont {G.~T.}\
  \bibnamefont {Fleming}}, \bibinfo {author} {\bibfnamefont {R.~C.}\
  \bibnamefont {Brower}}, \bibinfo {author} {\bibfnamefont {V.}~\bibnamefont
  {Ayyar}}, \bibinfo {author} {\bibfnamefont {E.~K.}\ \bibnamefont {Owen}},
  \bibinfo {author} {\bibfnamefont {T.~G.}\ \bibnamefont {Raben}},\ and\
  \bibinfo {author} {\bibfnamefont {C.-I.}\ \bibnamefont {Tan}},\ }\bibfield
  {title} {\bibinfo {title} {{Computing the Central Charge of the 3D Ising CFT
  Using Quantum Finite Elements}},\ }\href
  {https://doi.org/10.22323/1.430.0370} {\bibfield  {journal} {\bibinfo
  {journal} {PoS}\ }\textbf {\bibinfo {volume} {LATTICE2022}},\ \bibinfo
  {pages} {370} (\bibinfo {year} {2023})}\BibitemShut {NoStop}%
\bibitem [{\citenamefont {Han}\ \emph {et~al.}(2023)\citenamefont {Han},
  \citenamefont {Hu}, \citenamefont {Zhu},\ and\ \citenamefont
  {He}}]{Han:2023yyb}%
  \BibitemOpen
  \bibfield  {author} {\bibinfo {author} {\bibfnamefont {C.}~\bibnamefont
  {Han}}, \bibinfo {author} {\bibfnamefont {L.}~\bibnamefont {Hu}}, \bibinfo
  {author} {\bibfnamefont {W.}~\bibnamefont {Zhu}},\ and\ \bibinfo {author}
  {\bibfnamefont {Y.-C.}\ \bibnamefont {He}},\ }\bibfield  {title} {\bibinfo
  {title} {{Conformal four-point correlators of the 3D Ising transition via the
  quantum fuzzy sphere}},\ }\href@noop {} {\  (\bibinfo {year} {2023})},\
  \Eprint {https://arxiv.org/abs/2306.04681} {arXiv:2306.04681
  [cond-mat.stat-mech]} \BibitemShut {NoStop}%
\bibitem [{\citenamefont {Hu}\ \emph {et~al.}(2023)\citenamefont {Hu},
  \citenamefont {He},\ and\ \citenamefont {Zhu}}]{Hu:2023xak}%
  \BibitemOpen
  \bibfield  {author} {\bibinfo {author} {\bibfnamefont {L.}~\bibnamefont
  {Hu}}, \bibinfo {author} {\bibfnamefont {Y.-C.}\ \bibnamefont {He}},\ and\
  \bibinfo {author} {\bibfnamefont {W.}~\bibnamefont {Zhu}},\ }\bibfield
  {title} {\bibinfo {title} {Operator product expansion coefficients of the 3d
  ising criticality via quantum fuzzy spheres},\ }\href
  {https://doi.org/10.1103/PhysRevLett.131.031601} {\bibfield  {journal}
  {\bibinfo  {journal} {Phys. Rev. Lett.}\ }\textbf {\bibinfo {volume} {131}},\
  \bibinfo {pages} {031601} (\bibinfo {year} {2023})}\BibitemShut {NoStop}%
\bibitem [{\citenamefont {Di~Francesco}\ \emph {et~al.}(1997)\citenamefont
  {Di~Francesco}, \citenamefont {Mathieu},\ and\ \citenamefont
  {Senechal}}]{DiFrancesco1997}%
  \BibitemOpen
  \bibfield  {author} {\bibinfo {author} {\bibfnamefont {P.}~\bibnamefont
  {Di~Francesco}}, \bibinfo {author} {\bibfnamefont {P.}~\bibnamefont
  {Mathieu}},\ and\ \bibinfo {author} {\bibfnamefont {D.}~\bibnamefont
  {Senechal}},\ }\href {https://doi.org/10.1007/978-1-4612-2256-9} {\emph
  {\bibinfo {title} {Conformal Field Theory}}},\ Graduate Texts in Contemporary
  Physics\ (\bibinfo  {publisher} {Springer-Verlag},\ \bibinfo {address} {New
  York},\ \bibinfo {year} {1997})\BibitemShut {NoStop}%
\bibitem [{\citenamefont {Rychkov}(2016)}]{Rychkov:2016iqz}%
  \BibitemOpen
  \bibfield  {author} {\bibinfo {author} {\bibfnamefont {S.}~\bibnamefont
  {Rychkov}},\ }\bibfield  {title} {\bibinfo {title} {{EPFL Lectures on
  Conformal Field Theory in $D \ge 3$ Dimensions}},\ }\href@noop {} {\
  (\bibinfo {year} {2016})},\ \Eprint {https://arxiv.org/abs/1601.05000}
  {arXiv:1601.05000 [hep-th]} \BibitemShut {NoStop}%
\bibitem [{\citenamefont {Dolan}\ and\ \citenamefont
  {Osborn}(2001)}]{Dolan:2000ut}%
  \BibitemOpen
  \bibfield  {author} {\bibinfo {author} {\bibfnamefont {F.~A.}\ \bibnamefont
  {Dolan}}\ and\ \bibinfo {author} {\bibfnamefont {H.}~\bibnamefont {Osborn}},\
  }\bibfield  {title} {\bibinfo {title} {{Conformal four point functions and
  the operator product expansion}},\ }\href
  {https://doi.org/10.1016/S0550-3213(01)00013-X} {\bibfield  {journal}
  {\bibinfo  {journal} {Nucl. Phys. B}\ }\textbf {\bibinfo {volume} {599}},\
  \bibinfo {pages} {459} (\bibinfo {year} {2001})},\ \Eprint
  {https://arxiv.org/abs/hep-th/0011040} {arXiv:hep-th/0011040} \BibitemShut
  {NoStop}%
\bibitem [{\citenamefont {Dolan}\ and\ \citenamefont
  {Osborn}(2011)}]{Dolan:2011dv}%
  \BibitemOpen
  \bibfield  {author} {\bibinfo {author} {\bibfnamefont {F.~A.}\ \bibnamefont
  {Dolan}}\ and\ \bibinfo {author} {\bibfnamefont {H.}~\bibnamefont {Osborn}},\
  }\bibfield  {title} {\bibinfo {title} {{Conformal Partial Waves: Further
  Mathematical Results}},\ }\href@noop {} {\  (\bibinfo {year} {2011})},\
  \Eprint {https://arxiv.org/abs/1108.6194} {arXiv:1108.6194 [hep-th]}
  \BibitemShut {NoStop}%
\bibitem [{\citenamefont {Hogervorst}\ and\ \citenamefont
  {Rychkov}(2013)}]{Hogervorst2013}%
  \BibitemOpen
  \bibfield  {author} {\bibinfo {author} {\bibfnamefont {M.}~\bibnamefont
  {Hogervorst}}\ and\ \bibinfo {author} {\bibfnamefont {S.}~\bibnamefont
  {Rychkov}},\ }\bibfield  {title} {\bibinfo {title} {{Radial Coordinates for
  Conformal Blocks}},\ }\href {https://doi.org/10.1103/PhysRevD.87.106004}
  {\bibfield  {journal} {\bibinfo  {journal} {Phys. Rev. D}\ }\textbf {\bibinfo
  {volume} {87}},\ \bibinfo {pages} {106004} (\bibinfo {year} {2013})},\
  \Eprint {https://arxiv.org/abs/1303.1111} {arXiv:1303.1111 [hep-th]}
  \BibitemShut {NoStop}%
\bibitem [{\citenamefont {Costa}\ \emph {et~al.}(2016)\citenamefont {Costa},
  \citenamefont {Hansen}, \citenamefont {Penedones},\ and\ \citenamefont
  {Trevisani}}]{Costa2016}%
  \BibitemOpen
  \bibfield  {author} {\bibinfo {author} {\bibfnamefont {M.~S.}\ \bibnamefont
  {Costa}}, \bibinfo {author} {\bibfnamefont {T.}~\bibnamefont {Hansen}},
  \bibinfo {author} {\bibfnamefont {J.~a.}\ \bibnamefont {Penedones}},\ and\
  \bibinfo {author} {\bibfnamefont {E.}~\bibnamefont {Trevisani}},\ }\bibfield
  {title} {\bibinfo {title} {{Radial expansion for spinning conformal
  blocks}},\ }\href {https://doi.org/10.1007/JHEP07(2016)057} {\bibfield
  {journal} {\bibinfo  {journal} {JHEP}\ }\textbf {\bibinfo {volume} {07}},\
  \bibinfo {pages} {057}},\ \Eprint {https://arxiv.org/abs/1603.05552}
  {arXiv:1603.05552 [hep-th]} \BibitemShut {NoStop}%
\bibitem [{\citenamefont {Brower}\ and\ \citenamefont
  {Tamayo}(1989)}]{Brower:1989mt}%
  \BibitemOpen
  \bibfield  {author} {\bibinfo {author} {\bibfnamefont {R.~C.}\ \bibnamefont
  {Brower}}\ and\ \bibinfo {author} {\bibfnamefont {P.}~\bibnamefont
  {Tamayo}},\ }\bibfield  {title} {\bibinfo {title} {{Embedded Dynamics for
  $\phi^4$ Theory}},\ }\href {https://doi.org/10.1103/PhysRevLett.62.1087}
  {\bibfield  {journal} {\bibinfo  {journal} {Phys. Rev. Lett.}\ }\textbf
  {\bibinfo {volume} {62}},\ \bibinfo {pages} {1087} (\bibinfo {year}
  {1989})}\BibitemShut {NoStop}%
\bibitem [{\citenamefont {Metropolis}\ \emph {et~al.}(1953)\citenamefont
  {Metropolis}, \citenamefont {Rosenbluth}, \citenamefont {Rosenbluth},
  \citenamefont {Teller},\ and\ \citenamefont {Teller}}]{Metropolis:1953am}%
  \BibitemOpen
  \bibfield  {author} {\bibinfo {author} {\bibfnamefont {N.}~\bibnamefont
  {Metropolis}}, \bibinfo {author} {\bibfnamefont {A.~W.}\ \bibnamefont
  {Rosenbluth}}, \bibinfo {author} {\bibfnamefont {M.~N.}\ \bibnamefont
  {Rosenbluth}}, \bibinfo {author} {\bibfnamefont {A.~H.}\ \bibnamefont
  {Teller}},\ and\ \bibinfo {author} {\bibfnamefont {E.}~\bibnamefont
  {Teller}},\ }\bibfield  {title} {\bibinfo {title} {{Equation of state
  calculations by fast computing machines}},\ }\href
  {https://doi.org/10.1063/1.1699114} {\bibfield  {journal} {\bibinfo
  {journal} {J. Chem. Phys.}\ }\textbf {\bibinfo {volume} {21}},\ \bibinfo
  {pages} {1087} (\bibinfo {year} {1953})}\BibitemShut {NoStop}%
\bibitem [{\citenamefont {Adler}(1981)}]{Adler:1981sn}%
  \BibitemOpen
  \bibfield  {author} {\bibinfo {author} {\bibfnamefont {S.~L.}\ \bibnamefont
  {Adler}},\ }\bibfield  {title} {\bibinfo {title} {{An Overrelaxation Method
  for the Monte Carlo Evaluation of the Partition Function for Multiquadratic
  Actions}},\ }\href {https://doi.org/10.1103/PhysRevD.23.2901} {\bibfield
  {journal} {\bibinfo  {journal} {Phys. Rev. D}\ }\textbf {\bibinfo {volume}
  {23}},\ \bibinfo {pages} {2901} (\bibinfo {year} {1981})}\BibitemShut
  {NoStop}%
\bibitem [{\citenamefont {Whitmer}(1984)}]{Whitmer:1984he}%
  \BibitemOpen
  \bibfield  {author} {\bibinfo {author} {\bibfnamefont {C.}~\bibnamefont
  {Whitmer}},\ }\bibfield  {title} {\bibinfo {title} {{Overrelaxation methods
  for Monte Carlo simulations of quadratic and multiquadratic actions}},\
  }\href {https://doi.org/10.1103/PhysRevD.29.306} {\bibfield  {journal}
  {\bibinfo  {journal} {Phys. Rev. D}\ }\textbf {\bibinfo {volume} {29}},\
  \bibinfo {pages} {306} (\bibinfo {year} {1984})}\BibitemShut {NoStop}%
\bibitem [{\citenamefont {Brown}\ and\ \citenamefont
  {Woch}(1987)}]{Brown:1987rra}%
  \BibitemOpen
  \bibfield  {author} {\bibinfo {author} {\bibfnamefont {F.~R.}\ \bibnamefont
  {Brown}}\ and\ \bibinfo {author} {\bibfnamefont {T.~J.}\ \bibnamefont
  {Woch}},\ }\bibfield  {title} {\bibinfo {title} {{Overrelaxed Heat Bath and
  Metropolis Algorithms for Accelerating Pure Gauge Monte Carlo
  Calculations}},\ }\href {https://doi.org/10.1103/PhysRevLett.58.2394}
  {\bibfield  {journal} {\bibinfo  {journal} {Phys. Rev. Lett.}\ }\textbf
  {\bibinfo {volume} {58}},\ \bibinfo {pages} {2394} (\bibinfo {year}
  {1987})}\BibitemShut {NoStop}%
\bibitem [{\citenamefont {Creutz}(1987)}]{Creutz:1987xi}%
  \BibitemOpen
  \bibfield  {author} {\bibinfo {author} {\bibfnamefont {M.}~\bibnamefont
  {Creutz}},\ }\bibfield  {title} {\bibinfo {title} {{Overrelaxation and Monte
  Carlo Simulation}},\ }\href {https://doi.org/10.1103/PhysRevD.36.515}
  {\bibfield  {journal} {\bibinfo  {journal} {Phys. Rev. D}\ }\textbf {\bibinfo
  {volume} {36}},\ \bibinfo {pages} {515} (\bibinfo {year} {1987})}\BibitemShut
  {NoStop}%
\bibitem [{\citenamefont {Glück}\ \emph {et~al.}(2023)\citenamefont {Glück},
  \citenamefont {Ayyar}, \citenamefont {Brower}, \citenamefont {Fleming},
  \citenamefont {Owen}, \citenamefont {Raben},\ and\ \citenamefont
  {Tan}}]{Zenodo}%
  \BibitemOpen
  \bibfield  {author} {\bibinfo {author} {\bibfnamefont {A.-M.~E.}\
  \bibnamefont {Glück}}, \bibinfo {author} {\bibfnamefont {V.}~\bibnamefont
  {Ayyar}}, \bibinfo {author} {\bibfnamefont {R.~C.}\ \bibnamefont {Brower}},
  \bibinfo {author} {\bibfnamefont {G.~T.}\ \bibnamefont {Fleming}}, \bibinfo
  {author} {\bibfnamefont {E.~K.}\ \bibnamefont {Owen}}, \bibinfo {author}
  {\bibfnamefont {T.~G.}\ \bibnamefont {Raben}},\ and\ \bibinfo {author}
  {\bibfnamefont {C.-I.}\ \bibnamefont {Tan}},\ }\href
  {https://doi.org/10.5281/zenodo.10039534} {\bibinfo {title} {Dataset for
  "{T}he {O}perator {P}roduct {E}xpansion for {R}adial {L}attice {Q}uantization
  of 3{D} \texorpdfstring{$\phi^4$}{p4} {T}heory"}} (\bibinfo {year}
  {2023})\BibitemShut {NoStop}%
\bibitem [{\citenamefont {Jay}\ and\ \citenamefont {Neil}(2021)}]{Jay2020}%
  \BibitemOpen
  \bibfield  {author} {\bibinfo {author} {\bibfnamefont {W.~I.}\ \bibnamefont
  {Jay}}\ and\ \bibinfo {author} {\bibfnamefont {E.~T.}\ \bibnamefont {Neil}},\
  }\bibfield  {title} {\bibinfo {title} {{Bayesian model averaging for analysis
  of lattice field theory results}},\ }\href
  {https://doi.org/10.1103/PhysRevD.103.114502} {\bibfield  {journal} {\bibinfo
   {journal} {Phys. Rev. D}\ }\textbf {\bibinfo {volume} {103}},\ \bibinfo
  {pages} {114502} (\bibinfo {year} {2021})},\ \Eprint
  {https://arxiv.org/abs/2008.01069} {arXiv:2008.01069 [stat.ME]} \BibitemShut
  {NoStop}%
\bibitem [{\citenamefont {Byrd}\ \emph {et~al.}(1995)\citenamefont {Byrd},
  \citenamefont {Lu}, \citenamefont {Nocedal},\ and\ \citenamefont
  {Zhu}}]{Byrd1995}%
  \BibitemOpen
  \bibfield  {author} {\bibinfo {author} {\bibfnamefont {R.~H.}\ \bibnamefont
  {Byrd}}, \bibinfo {author} {\bibfnamefont {P.}~\bibnamefont {Lu}}, \bibinfo
  {author} {\bibfnamefont {J.}~\bibnamefont {Nocedal}},\ and\ \bibinfo {author}
  {\bibfnamefont {C.}~\bibnamefont {Zhu}},\ }\bibfield  {title} {\bibinfo
  {title} {A limited memory algorithm for bound constrained optimization},\
  }\href@noop {} {\bibfield  {journal} {\bibinfo  {journal} {SIAM Journal on
  Scientific Computing}\ }\textbf {\bibinfo {volume} {16}},\ \bibinfo {pages}
  {1190} (\bibinfo {year} {1995})}\BibitemShut {NoStop}%
\bibitem [{Note1()}]{Note1}%
  \BibitemOpen
  \bibinfo {note} {For \(L=36\) some of the most probable fits had unphysical
  values of \(\Delta _{\epsilon '} < \Delta _T\), which we
  discarded.}\BibitemShut {Stop}%
\bibitem [{\citenamefont {Strang}\ and\ \citenamefont
  {Fix}(2008)}]{StrangFix200805}%
  \BibitemOpen
  \bibfield  {author} {\bibinfo {author} {\bibfnamefont {G.}~\bibnamefont
  {Strang}}\ and\ \bibinfo {author} {\bibfnamefont {G.}~\bibnamefont {Fix}},\
  }\href {http://amazon.com/o/ASIN/0980232708/} {\emph {\bibinfo {title} {An
  Analysis of the Finite Element Method 2nd Edition}}},\ \bibinfo {edition}
  {2nd}\ ed.\ (\bibinfo  {publisher} {Wellesley-Cambridge},\ \bibinfo {year}
  {2008})\BibitemShut {NoStop}%
\bibitem [{\citenamefont {{Desbrun}}\ \emph {et~al.}(2005)\citenamefont
  {{Desbrun}}, \citenamefont {{Hirani}}, \citenamefont {{Leok}},\ and\
  \citenamefont {{Marsden}}}]{2005math8341D}%
  \BibitemOpen
  \bibfield  {author} {\bibinfo {author} {\bibfnamefont {M.}~\bibnamefont
  {{Desbrun}}}, \bibinfo {author} {\bibfnamefont {A.~N.}\ \bibnamefont
  {{Hirani}}}, \bibinfo {author} {\bibfnamefont {M.}~\bibnamefont {{Leok}}},\
  and\ \bibinfo {author} {\bibfnamefont {J.~E.}\ \bibnamefont {{Marsden}}},\
  }\bibfield  {title} {\bibinfo {title} {{Discrete Exterior Calculus}},\
  }\href@noop {} {\bibfield  {journal} {\bibinfo  {journal} {ArXiv Mathematics
  e-prints}\ } (\bibinfo {year} {2005})},\ \Eprint
  {https://arxiv.org/abs/math/0508341} {math/0508341} \BibitemShut {NoStop}%
\bibitem [{\citenamefont {Brower}\ and\ \citenamefont
  {Owen}(2023)}]{Brower_2023}%
  \BibitemOpen
  \bibfield  {author} {\bibinfo {author} {\bibfnamefont {R.~C.}\ \bibnamefont
  {Brower}}\ and\ \bibinfo {author} {\bibfnamefont {E.~K.}\ \bibnamefont
  {Owen}},\ }\bibfield  {title} {\bibinfo {title} {Ising model on the affine
  plane},\ }\bibfield  {journal} {\bibinfo  {journal} {Physical Review D}\
  }\textbf {\bibinfo {volume} {108}},\ \href
  {https://doi.org/10.1103/physrevd.108.014511} {10.1103/physrevd.108.014511}
  (\bibinfo {year} {2023})\BibitemShut {NoStop}%
\bibitem [{\citenamefont {Blote}\ \emph {et~al.}(1995)\citenamefont {Blote},
  \citenamefont {Luijten},\ and\ \citenamefont {Heringa}}]{Blote:1995zik}%
  \BibitemOpen
  \bibfield  {author} {\bibinfo {author} {\bibfnamefont {H.~W.~J.}\
  \bibnamefont {Blote}}, \bibinfo {author} {\bibfnamefont {E.}~\bibnamefont
  {Luijten}},\ and\ \bibinfo {author} {\bibfnamefont {J.~R.}\ \bibnamefont
  {Heringa}},\ }\bibfield  {title} {\bibinfo {title} {{Ising universality in
  three dimensions: a Monte Carlo study}},\ }\href
  {https://doi.org/10.1088/0305-4470/28/22/007} {\bibfield  {journal} {\bibinfo
   {journal} {J. Phys. A}\ }\textbf {\bibinfo {volume} {28}},\ \bibinfo {pages}
  {6289} (\bibinfo {year} {1995})},\ \Eprint
  {https://arxiv.org/abs/cond-mat/9509016} {arXiv:cond-mat/9509016}
  \BibitemShut {NoStop}%
\bibitem [{\citenamefont {Hasenbusch}\ \emph {et~al.}(1999)\citenamefont
  {Hasenbusch}, \citenamefont {Pinn},\ and\ \citenamefont
  {Vinti}}]{Hasenbusch:1998gh}%
  \BibitemOpen
  \bibfield  {author} {\bibinfo {author} {\bibfnamefont {M.}~\bibnamefont
  {Hasenbusch}}, \bibinfo {author} {\bibfnamefont {K.}~\bibnamefont {Pinn}},\
  and\ \bibinfo {author} {\bibfnamefont {S.}~\bibnamefont {Vinti}},\ }\bibfield
   {title} {\bibinfo {title} {{Critical exponents of the three-dimensional
  Ising universality class from finite-size scaling with standard and improved
  actions}},\ }\href {https://doi.org/10.1103/PhysRevB.59.11471} {\bibfield
  {journal} {\bibinfo  {journal} {Phys. Rev. B}\ }\textbf {\bibinfo {volume}
  {59}},\ \bibinfo {pages} {11471} (\bibinfo {year} {1999})},\ \Eprint
  {https://arxiv.org/abs/hep-lat/9806012} {arXiv:hep-lat/9806012} \BibitemShut
  {NoStop}%
\end{thebibliography}%

\end{document}